\documentclass[sigconf]{acmart}
\usepackage{multirow}
\usepackage{caption} 
\usepackage{subcaption}
\usepackage{algorithm}
\usepackage{natbib}
\usepackage{algpseudocode}
\usepackage[english]{babel}
\usepackage{amsthm}
\theoremstyle{definition}
\newtheorem{definition}{Definition}[subsection]
\AtBeginDocument{%
  }

\setcopyright{acmlicensed}
\copyrightyear{2018}
\acmYear{2018}
\acmDOI{XXXXXXX.XXXXXXX}

\acmConference[Conference acronym 'XX]{Make sure to enter the correct
  conference title from your rights confirmation emai}{June 03--05,
  2018}{Woodstock, NY}

\acmISBN{978-1-4503-XXXX-X/18/06}

\begin{document}

\newcommand*{\ouralgo}{\text{RS-FairFRS}}
\title{Towards Fairness in Provably Communication-Efficient Federated Recommender Systems}

\author{Kirandeep Kaur}
\affiliation{%
  \institution{Indian Institute of Technology}
  \city{Ropar}
  \country{India}}
\email{staff.kirandeep.kaur@iitrpr.ac.in}

\author{Sujit Gujar}
\affiliation{%
  \institution{International Institute of Technology}
  \city{Hyderabad}
  \country{India}}
\email{sujit.gujar@iiit.ac.in}

\author{Shweta Jain}
\affiliation{%
  \institution{Indian Institute of Technology}
  \city{Ropar}
  \country{India}}
\email{shweta.jain@iitrpr.ac.in}

\begin{abstract}
To reduce the communication overhead caused by parallel training of multiple clients, various federated learning (FL) techniques use random client sampling. Nonetheless, ensuring the efficacy of random sampling and determining the optimal number of clients to sample in federated recommender systems (FRSs) remains challenging due to the isolated nature of each user as a separate client. This challenge is exacerbated in models where public and private features can be separated, and FL allows communication of only public features (item gradients). In this study, we establish sample complexity bounds that dictate the ideal number of clients required for improved communication efficiency and retained accuracy in such models. In line with our theoretical findings, we empirically demonstrate that our \ouralgo~reduces communication cost ($\approx47\%$). Second, we demonstrate the presence of class imbalance among clients that raises a substantial equity concern for FRSs. Unlike centralized machine learning, clients in FRS can not share raw data, including sensitive attributes. For this, we introduce \ouralgo~, first fairness under unawareness FRS built upon \emph{random sampling} based FRS. While random sampling improves communication efficiency,  we propose a novel two-phase \emph{dual-fair update} technique to achieve fairness without revealing protected attributes of \emph{active} clients participating in training. Our results on real-world datasets and different sensitive features illustrate a significant reduction in demographic bias ($\approx40\%$), offering a promising path to achieving fairness and communication efficiency in FRSs without compromising the overall accuracy of FRS.
\end{abstract}




%%
%% Keywords. The author(s) should pick words that accurately describe
%% the work being presented. Separate the keywords with commas.

\received{20 February 2007}
\received[revised]{12 March 2009}
\received[accepted]{5 June 2009}

\maketitle

%%%%%%%%%%%%%%%%%%%%%%%%%%%%%%%%%%%%%%%%%%%%%
\section{Introduction}
\label{sec:intro}
\emph{Recommender systems} (RSs) adopted by various online platforms use raw user data for training at the central server. Utilizing such sensitive data continues to spark ethical and privacy concerns. Regulations such as the General Data Protection Regulation (GDPR),\footnote{https://gdpr-info.eu/}, Personal Information Protection Law of the People’s Republic of China (PIPL)\footnote{https://personalinformationprotectionlaw.com/} and the California Consumer Privacy Act (CCPA)\footnote{https://oag.ca.gov/privacy/ccpa} have been established to address growing consumer apprehension and protect privacy rights of end users.

\emph{Federated learning (FL)}~\cite{mcmahan2017communicationefficient} has emerged as a potential solution for securing user's privacy. FL is a decentralized approach to train machine learning models that requires no exchange of data between client devices and global servers.  In the realm of RSs, FL has been adopted to generate personalized recommendations~\cite{4}, context-aware recommendations~\cite{ali2021federated}, online recommendations~\cite{zhou2012federated} and sequential recommendations~\cite{han2021deeprec}. 
The FedRec~\cite{FedRec}, an FRS that trains local clients using \emph{matrix factorization} (MF) and sends updates to the server without sharing raw user data. MF, a simple embedding model, maps users and items onto a shared latent space, generating latent factors in the form of \emph{user vectors} and \emph{item vectors}, whose dot product generates predictions; only item updates are shared with the server. Subsequently, the server aggregates these local updates to generate a global model and disseminate it to all clients.

MF, a form of collaborative filtering~\cite{3},
remains a cornerstone technique in many online platforms~\cite{India_2019,Natashia_2023a}. Despite the availability of deep models,~\citet{rendle2020neural} shows that a simple MF dot product outperforms a multilayer perceptron (MLP)\footnote{A multilayer perceptron (MLP) is a type of feedforward artificial neural network that consists of multiple layers of neurons used in machine learning and deep learning techniques}, making the latter impractical for recommendation. Parallel training, communication of gradients, and the use of MLP can abruptly boost communication and computation costs in FRSs, affecting their practicality. Thus, this paper uses FedRec~\cite{FedRec} because of its simplicity and efficiency.

While \textit{random sampling of clients } is widely used to improve communication efficiency in  FL models, proving its efficiency in an MF-based FRS is challenging as clients communicate only item updates with the server. \textbf{This paper is the first to establish a theoretical bound on the optimal fraction of clients to be sampled for improved communication costs yet preserved model accuracy. }Collaborative filtering algorithms use a matching-based paradigm~\cite{mf,adomavicius2005toward,chen2021neural,jannach2010recommender} by presuming similarities in client preferences for projecting users and items onto a shared embedding space. This forms an inherent clustering among clients. We leverage this widely used assumption to demonstrate that random sampling sufficiently captures representation from each such cluster in an FRS. We show that the ratings predicted after sampling closely align (with high probability) with the ratings generated with the participation of all clients.

Additionally, we address the issue of demographic bias in such systems. For example, LinkedIn's job recommendation model was recently found to be discriminatory against women since men seek jobs more frequently~\cite{Wall_2021}.  Further, Google advertisement ~\cite{Bradford_2022} was shown to display 

better jobs for men than women. Such biased treatment perpetuates employment discrimination and fortifies social stereotypes.

Fairness in an FRS, a critical issue, remains insufficiently explored and addressed. We first demonstrate that FedRec offers superior recommendations to a specific group of users. Fairness approaches in centralized RSs necessitate disclosure of sensitive attributes~\cite{in1, post2}, but users in federated scenarios are not required to share sensitive features. Hence, we tackle the issue of demographic bias in FRS without mandating \emph{active} users to reveal their sensitive attributes. In federated classification,~\citet{gifairfl, Kanaparthy2023} alleviate group bias by training locally with a fair global model. Consequently, relying solely on fair global models can perpetuate bias due to personalized local training via MF, potentially resulting in an overall biased model.

Thus, \textbf{we introduce \ouralgo~a novel dual-fair vector update method coupled with a two-phase fairness training approach. }The server aggregates received item gradients and trains in the initial phase to achieve a fair global model using uses \emph{FairMF}. Despite a fair global model, local client updates could significantly skew the recommendations. Therefore, clients minimize local loss and the difference between learnt item vectors and a fair global model in the subsequent phase for improved fairness. Through this approach, we address the pivotal challenge of \emph{demographic bias} in FRS. It should be noted that we bound the optimal number of clients only for improving \emph{communication efficiency} and ensure fairness using dual-fair vector updates in such communication efficient FRS.
Our main contributions are: 
 
\begin{enumerate}
  \item Our work provides \textit{sample complexity bounds} on the fraction of clients required to preserve accuracy in FRS (Theorem~\ref{thm:main}). Empirically, we show that sampling these many clients ($35\%$) reduces communication costs ($\approx47\%$). 
  
  \item We propose \emph{dual-fairness vector update} technique that updates item vectors in two phases using \emph{FairMF} (for fair global model) and \emph{FO-Client Batch} (for fairness oriented local training) to mitigate group bias in random sampling based FRS.
  
  \item\ouralgo,~a novel FRS, enhances communication efficiency by selecting an optimal number of clients and significantly improves fairness by using the dual-fair updates.
  
  \item Experiments on the widely used ML1M and ML100K datasets, with diverse demographics (age and gender) demonstrate that \ouralgo~successfully reduces demographic bias ($\approx40\%$) while simultaneously improving accuracy.
\end{enumerate}

%%%%%%%%%%%%%%%%%%%%%%%%%%%%%
\section{Related Work}
\label{sec:related_work}
%%%%%%%%%%%%%%%%%%%%%%%%%%%%%
\subsection{Client Sampling in FL } \textit{Partial participation}~\cite{serverside} in FL
addresses the generalization issues, diminishing returns, training failures, and fairness concerns that arise when using large cohorts for training~\cite{largecohort}. Recent works~\cite{lai2021oort,amiri2021convergence,nguyen2020fast,cho2022towards,li2022pyramidfl,zhou2022you,sultana2022eiffel,luo2022tackling} exploit data and system heterogeneity to design various client selection strategies.~\citet{nishio2019client} and~\citet{xu2020client} sample as many clients as possible based on the given time and bandwidth, respectively.~\citet{chen2022optimal} design a learn-to-select system using a norm-based client sampling strategy, and~\citet{ribero2022federated} select clients based on their availability. Other techniques for client sampling include a greedy strategy to select clients representing the entire population~\cite{diverse}, sampling clients based on the norm of update~\cite{optimalclientsampling}, and clustered sampling to reduce variance~\cite{clusteredsampling}. A theoretical exploration by~\citet{generaltheory} investigates the impact of sampling on convergence, and~\citet{clientselection} provides an optimization convergence study for biased client selection techniques. 

These studies primarily concentrate on federated classification;~\citet{two,one} employ random sampling in FRSs. While Google's Gboard samples $100$ clients for generating keyboard query suggestions~\cite{10197174},~\citet{mcmahan2017communicationefficient} show that additional clients improve the convergence in FL. We close the research gap by determining the ideal fraction of clients to sample during communication rounds in FRS to maintain overall performance. 
\subsection{Fairness in Federated Recommendations } Mitigating bias in centralized recommendations can be trivial since raw user data is easily available at the server. Many past works on group fairness in RSs ~\cite{gf1,gf2,post1,gf4,beyondparityin3} require access to sensitive attributes of every user to quantify and mitigate bias at the group level.  Techniques that do not require sensitive attributes~\cite{gf3,deepfair} learn disparity in data while training to produce unbiased recommendations. However, in FRSs, most users might share only gradients with the server, complicating efforts to mitigate bias.

Fairness in federated recommendations is an under-investigated area. Different fairness aspects are considered in recent works, including Cali3F~\cite{four}, which aims to obtain a uniform NDCG score across clients; RF2~\cite{five}, which models the inter-dependence of data and system heterogeneity for fair recommendations; and ~\citet{six}, which designs adaptive techniques to capture data and performance imbalance among different clients. None of this work tackles demographic bias in an FRS. Closest to our work,~\citet{agrawal2023prejudice} mitigate group bias in the top-k recommendations system by using proxies (S=1/0) for binary protective groups, which can be interpreted at the server and result in privacy breaches. To the best of our knowledge, we are first to improve demographic bias in FRS when access to sensitive attributes of clients participating in communication round (\emph{active clients}) is unavailable. \ouralgo\ operates on a score prediction-based recommendation system without sharing any sensitive information.

We acknowledge various techniques proposed in federated classification settings for improving group fairness~\cite{minimax,Kanaparthy2023,gifairfl,fair_fed_virgina}.  These works cannot be applied directly to an FRS, where each user acts as one client; in classification schemes, one client can have data from multiple users. Some of these works require communication of sensitive attributes to the server, which increases communication overhead and affect the privacy of active users. We demonstrate that the availability of small amounts of seed data on the server can solve the issue of demographic fairness under unawareness in FRSs by protecting the privacy of \emph{active} users. Below, we present works that use server learning in federated scenarios to support our assumption.
 
 \subsection{FL with Server Learning }While most past works in FL consider the server as an aggregator,~\citet{serverlearning} posits that the server has access to a small amount of training data; e.g., automated vehicles require the identification of certain road objects. FSL, proposed by~\citet{serverlearning}, shows that if the server is given access to minimal training data, the convergence rate improves substantially, even at the onset of learning. Other works like~\citet{fedDUAP} use insensitive data on the server along with dynamic updates and adaptive pruning to reduce computation cost. ~\citet{Kanaparthy2023} use server-side information for a demographically fair face attribute classification task, and~\citet{personal} propose FedSIM, which uses server-side data to boost the estimation of meta-gradients and improve personalized performance. Data sharing~\cite{zhu2021federated} in  FL allows the server to hold a globally shared dataset used to warm up the server through the training on this dataset for handling non-IID data. While only a small amount of data on the server has been shown to mitigate various challenges in FL classification, \citet{wu2022federated} develop the first federated graph neural network framework for generating personalized recommendations in a privacy-preserving manner by using server learning. 

\section{Preliminaries}
\label{sec:prelim}
\subsection{Matrix Factorization Based FedRec }
~\citet{FedRec} adopts a federated learning approach to generate recommendations by using MF for local training. We acknowledge the recent shift towards utilizing learned similarity, such as employing a \textit{multilayer perceptron} (MLP) instead of the dot product used by traditional MF, an approach known as \textit{neural collaborative filtering} (NCF)~\cite{he2017neural}. However, its usage in federated scenarios can increase communication and computation costs, as suggested by~\citet{rendle2020neural}, who demonstrate that, with appropriate hyperparameter selection, a simple dot product often significantly outperforms the proposed learned similarities. Thus, to make deployment of the FRS more pragmatic, we recognize the importance of using MF for local training, as was done by~\citet{FedRec}.

Given $n$ users (or clients) and $m$ items in a typical FRS with explicit feedback, each user $u \in \{1,2,\ldots,n\}$ has a rating vector $[r_{ui}]_{i=1}^m$ that depicts the rated items $i \in \{1,2,\ldots,m\}$. We denote the true and predicted ratings given by the user $u$ for an item $i$ using $r_{ui}$ and $\hat{r}_{ui}$, respectively. $p_{ui} \in \{0,1\}$ serves as an indicator variable that shows if the item is observed $(r_{ui} = 1)$ or not $(r_{ui} = 0)$. An MF-based algorithm works by generating two matrices, $U\in \mathbb{R}^{n \times k}$ and $V \in \mathbb{R} ^ {m \times k}$, such that each client $u$ is associated with a vector $U_u \in \mathbb{R}^{k}$ (\emph{user vector}) and each item $i$ is associated with $V_i \in \mathbb{R}^{k}$(\emph{item vector}). The predicted rating of the $i^{th}$ item by the $u^{th}$ user can be computed as $\hat{r}_{ui}= U_u. V_{i}^{T}$. 

MF-based algorithms, then,  aim to learn these user vectors $U_u$ for each user $u$ and item vectors $V_i$ for each item $i$ and  minimize the following regularized loss function: 
\begin{equation}
   \mathcal{L^{MF}} = \sum_{u\in [n]}\sum_{i\in [m]} p_{ui}(r_{ui} - U_{u}.V_{i}^{T})^2 + \lambda^r(\mid\mid V_{i}\mid\mid^2 +\mid\mid U_{u}\mid\mid^2).
    \label{eq:MF_optimfactorization_func}
\end{equation}
The clients use $\mathcal{L^{MF}}$ for local training in FedRec. For improved privacy guarantees,  FedRec uses two techniques, \emph{user averaging} (UA) and \emph{hybrid filling} (HF), to add noise to local data. UA samples some unrated items and assigns them a virtual rating $r^{'}_{ui}$; HF replaces the virtual rating with the predicted rating after some time. To improve privacy,  gradients for the truly and virtually rated items are sent to the server. The normalized gradient of the user for true and virtually rated items $I_{u}^{'} \cup I_{u}$ is given as: 
\begin{equation}
 \hspace*{-4mm} \nabla U_u = \frac{\sum\limits_{i \in I^{'}_{u}\cup I_u} [(U_u V_i^T - p_{ui}r_{ui} - (1-p_{ui})r^{'}_{ui})V_i + \lambda^r U_u] }{|I^{'}_u \cup I_u|.}
\label{eq:FedRec_clientgradr}
\end{equation}
Each local client sends only $\nabla V (u,i)$ to the server, calculated as: 
\begin{align}
\nabla V (u,i) = \begin{cases}
(U_u V_i^T - r_{ui})U_u + \lambda^r V_i, p_{ui} = 1\\ 
(U_u V_i^T - r^{'}_{ui})U_u + \lambda^r V_i, p_{ui} = 0.
\end{cases}
\label{eq: fedrec_aggregate}
\end{align}
 The server then evaluates the aggregated gradient for each item $i$, as follows: 
\begin{equation}
\nabla V_i = \frac{1}{|U_i^{'} \cup U_i |}\sum\limits_{u \in U_i^{' }\cup U_i } \nabla V (u,i).
\label{eq:FedRec_Itemvector}
\end{equation}
Here, $U_{i}^{'}$ and $U_i$ denote the set of clients who rated an item $i$ virtually and truly, respectively. The server communicates a global model to all clients for local training. While random sampling improves communication efficiency in FL, we build upon the notion of fairness in the FRS, as discussed below.

\subsection{Group-Fairness Notion }
~\citet{pedreschi2009measuring} posits that achieving group fairness necessitates treating disadvantaged and advantaged groups in a similar manner. Many real-world datasets have class imbalance issues because users belonging to one attribute are usually under-represented relative to other attributes. Due to this imbalance, a model might not perform well on the under-represented community because of insufficient data. Thus, we identify advantaged and disadvantaged groups using the following definition:
\begin{definition}[%\emph{Group Stratification} 
]\emph{Let $U$ be a set of $n$ users, and $G$ be a binary attribute partitioning $U$ into two disjoint sets, $g$ and $\neg g'$, such that $g \cap \neg g = \emptyset$ and $|g| > | \neg g|.$ Given a recommendation model $\theta$, let $\mathcal{L}$ denotes the accuracy metric\footnote{It is important to note that our recommendation model aims to predict scores using a point-wise loss function. Thus, we use accuracy as the performance metric. In other scenarios where the task involves ranking or generating top-k recommendations, various metrics like NDCG, F1 score, precision and recall can be used.}. If }
\[
  \mathcal{L}(g) < \mathcal{L}(\neg g),
\]
\emph{then $g$ is the advantaged group and $\neg g$ is formally recognized as the disadvantaged group within the set $U$. Here, $|g|$ denotes the cardinality of group $g$ and $|\neg g|$ represents the cardinality of group $\neg g$.}
\end{definition}

Existing group fairness notions in an RS---namely, value unfairness, absolute unfairness, and non-parity fairness~\cite{beyondparityin3}---are used in centralized RSs to measure fairness. Such metrics evaluate the loss on each item based on the difference in average loss on a specific item $i$ for \emph{advantaged} and \emph{disadvantaged} groups. However, when only item gradients are communicated with the server, it becomes infeasible to evaluate the number of advantaged/disadvantaged users who rated an item. Thus, unlike the preceding notions, we use \emph{demographic accuracy parity} ($\mathcal{L}^{dap}$), which  normalizes the sum of squared loss over all items rated by a user: 
 
%\begin{tiny}
    \begin{equation}
        \frac{1}{|g|} \sum_{u \in g} \frac{1}{|I_u|} \sum_{I \in I_{u}}
        ( \hat{r}_{ui} - r_{ui})^2 - \frac{1}{|\neg g|} \sum_{u \in \neg g} \frac{1}{|I_u|} \sum_{I \in I_{u}}
        ( \hat{r}_{ui} - r_{ui})^2
        \label{eq:d_bias}.
    \end{equation}
%\end{tiny}
Here, $g$ and $\neg g$ denote the disadvantaged and advantaged groups, respectively. $I_u$ is the set of items rated by $u$, and $\hat{r}_{ui}$, $r_{ui}$ depict the predicted and true ratings, respectively.

\begin{definition}[]
\emph{Let $\theta^f$  and $\widehat{\theta}^f$ denote two federated recommendation models. We say that $\theta^f$ is a demographically fairer model if}
\[
\mathcal{L}^{dap}(\theta^f) < \mathcal{L}^{dap}(\widehat{\theta}^f).
\]
\end{definition}
To protect the privacy of active users who participate in federated training, we use the availability of seed data on the server to generate a fair global model. Importantly, users in the seed data do not participate in training. We illustrate our assumption by considering public availability of various datasets like the Netflix challenge. Big tech companies like Netflix can use such prior data as seed. $\mathcal{L}^{dap}$ can then be used in the objective function of MF to train seed data for fairness at the server. Further, local clients train towards a fair global model, obtaining fair recommendations locally. We now describe our proposed methodology in detail.
%__________________________________
\begin{figure}[t!]
     \centering
     \begin{subfigure}[b]{0.22\textwidth}
         \centering
         \includegraphics[width=\textwidth]{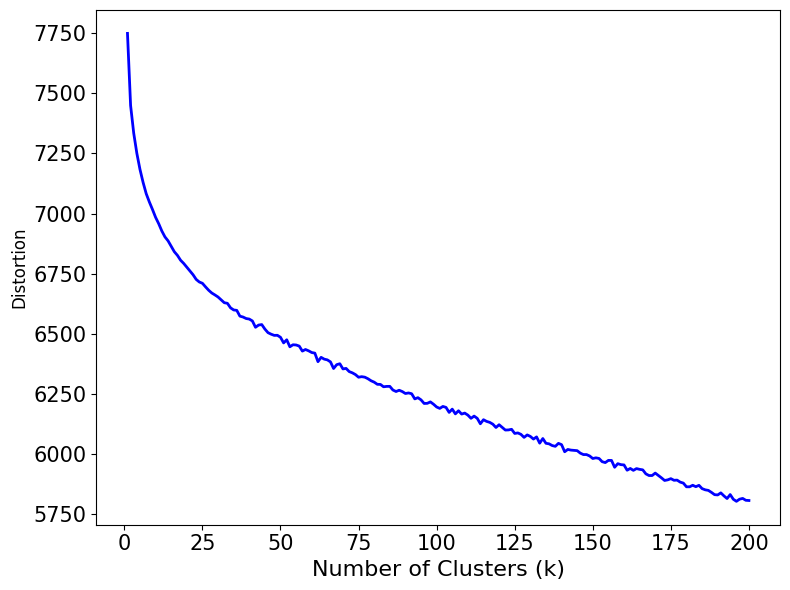}
         \caption{ML1M}
         \label{sfig:a1}
     \end{subfigure}
     %\hfill
     \begin{subfigure}[b]{0.22\textwidth}
         \centering
          \includegraphics[width=\textwidth]{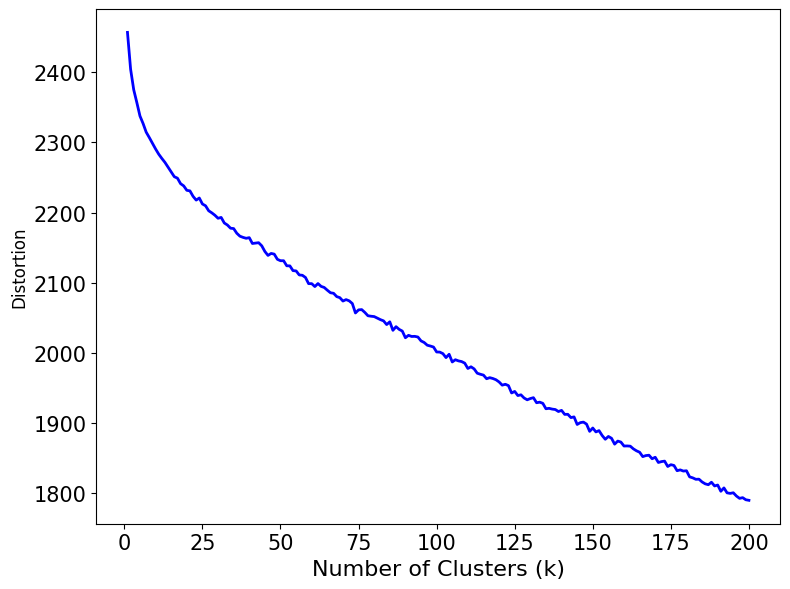}
         \caption{ML100k}
         \label{sfig:b1}
     \end{subfigure}
    \caption{Ideal number of clusters in both datasets.}
    \label{fig:cluster_plot1}
\end{figure}

%%%%%%%%%%%%%%%%%%%%%%%%%%%%%%%%%%%%%%
\section{Proposed Methodology}
\label{sec:ourapproach}
%%%%%%%%%%%%%%%%%%%%%%%%%%%%%%%%%%%%%%
Each communication round in an FRS requires the communication of item gradients of all participating clients to the server. This leads to sending a total of $n\times m \times k$ parameters, imposing a huge communication cost. To reduce this cost, we randomly sample clients in each communication round who communicate $\nabla V(u, i)$ with the server, as was done by~\citet{largecohort,generaltheory,clientselection}. Our theoretical contribution addresses an important question of how many clients we need to sample in each communication round to maintain the model's overall accuracy. 
%%%%%%%%%%%%%%%%%%%%%%%%%%%%%%%%%%%%%%%%
\subsection{Random Sampling without Replacement}
\label{sec:random-sampling}
%%%%%%%%%%%%%%%%%%%%%%%%%%%%%%%%%%%%%%%%

Models based on collaborative filtering~\cite{mf,gupta2020movie} work by identifying latent similarities among users based on their rating patterns%\footnote{We experimentally validated that the item vectors do form clusters.}. 
. We use this latent similarity among users to assume the existence of \textit{latent clusters} such that user data within the cluster remains independent and identically distributed. 
%We aim to utilize this homogeneity within the same clusters without the actual knowledge of the $K$ clusters and the users of each cluster to bound the ideal fraction of clients to be sampled. Note that the relation of this assumption is used only to provide the sample-bound complexity without addressing fairness so far. 

Let $C^{\tau}$ represent the set of clients selected after randomly sampling $\tau$ fraction of clients and let there exist some $K$ clusters into which the total number of clients can be grouped. Lemma~\ref{Lemma:1} shows that when a certain number of clients are sampled randomly 
at uniform, they represent each cluster equally to ensure that this sampled set is sufficient to represent the entire population~\cite{diverse}. 
%%%%%%%%% LEMMA1 %%%%%%%%%%%
\begin{lemma}
\label{Lemma:1}
%%%%%%%%% LEMMA1 %%%%%%%%%%%
Suppose $n$ clients are uniformly distributed amongst $K$ clusters. Then, a subset $C^\tau \subseteq [n]$ sampled uniformly at random (without replacement) will contain an approximately equal number of clients from each cluster, i.e., 

\[\mathbb{P}\left(\left|\sum_i X_i^j - \frac{|C^\tau|}{K}\right| \ge \epsilon\right)\] is very low. 
\end{lemma}
Here, $X_i^j \in \{0,1\}$ denotes the random variable taking the value $1$ when the  $i^{th}$ sample belongs to cluster $j$, and $0$ otherwise. Note that Hoeffding's bound gives
$\mathbb{P}\left(|\sum_i X_i^j - \frac{|C^\tau|}{K}| \ge \epsilon\right) \le 2 \exp \left\{\frac{-2\epsilon^2}{|C^\tau|}\right\}
$.  Substituting $|C^{\tau}| = 2100$, i.e., $35\%$ of the total number of clients ($n=6000$ in MovieLense dataset), we get this probability to be $0.62$ at $K=10$. This might be a loose bound, but the probability is very high, which is evident from the following experiments.
%some basic experiments in the Appendix.

Each user participating in one communication round generates user and item vectors. FedRec allows the communication of only item vectors. Hence, sampling one user means sampling the item vectors corresponding to that user. We cluster these item vectors using K-means clustering~\cite{bock2007clustering}. Figure \ref{fig:cluster_plot1} shows the elbow curve obtained after plotting clustering loss \emph{Inertia}~\cite{cc}, corresponding to the number of clusters $K$. Inertia (or distortion), one of the most popular metrics for evaluating clustering results, is calculated as the sum of distances to each cluster centre. A high value of total distance implies that the points are far from each other, a clear indication that they are less similar to each other. For both datasets, we get $K=20$ as the ideal number of clusters. 

%__________________________________
\begin{figure*}[ht!]
     \centering
     \begin{subfigure}[b]{0.23\textwidth}
         \centering
         \includegraphics[width=\textwidth]{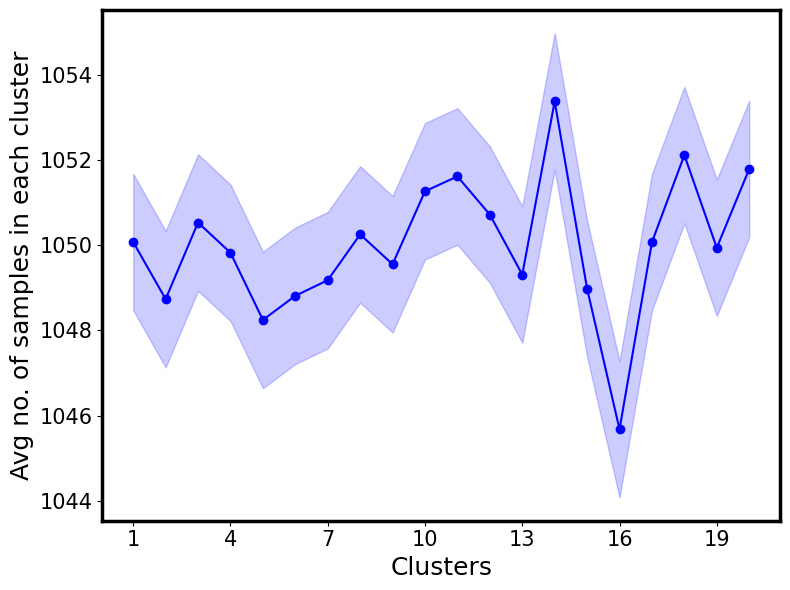}
         \caption{}
         \label{sfig:a2}
     \end{subfigure}
     %\hfill
    \hspace{0.2mm}
     \begin{subfigure}[b]{0.23\textwidth}
         \centering
       \includegraphics[width=\textwidth]{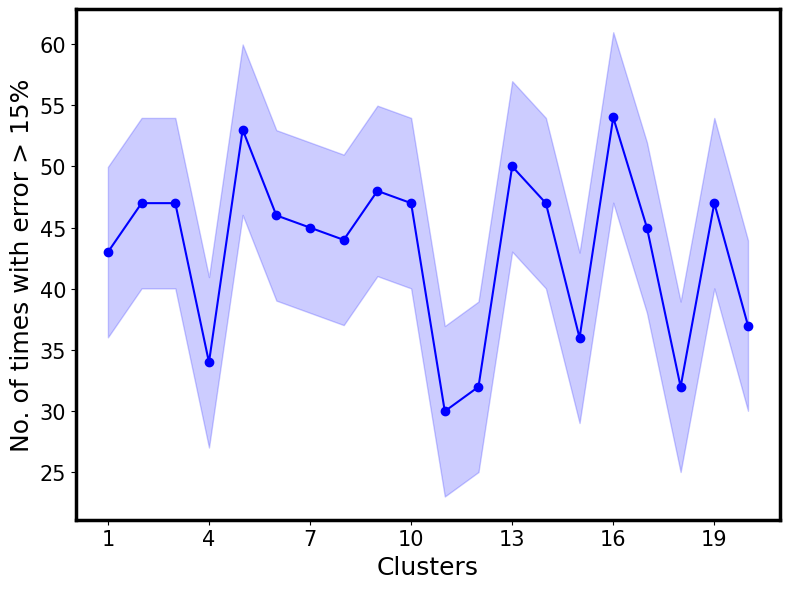}
         \caption{}
         \label{sfig:b2}
     \end{subfigure}
     %\hfill
    \hspace{0.2mm}
     \begin{subfigure}[b]{0.23\textwidth}
         \centering         \includegraphics[width=\textwidth]{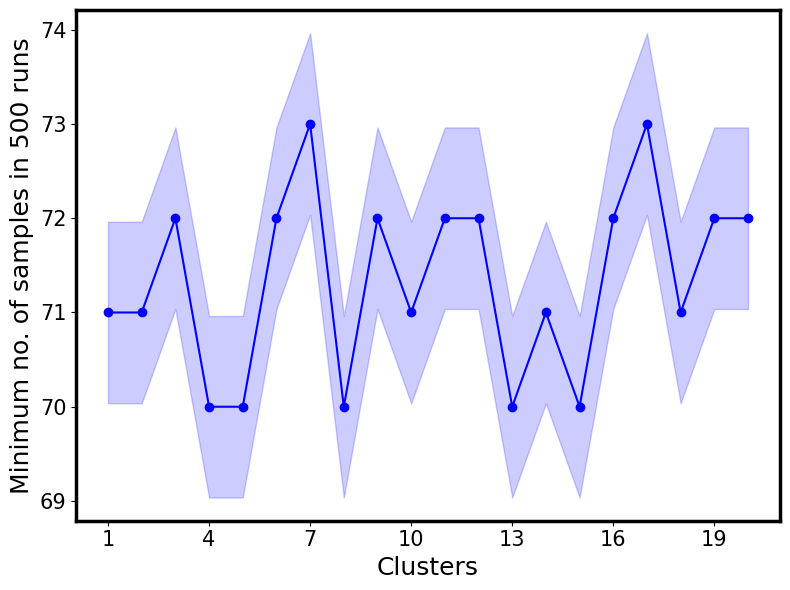}
         \caption{}
         \label{sfig:c2}
     \end{subfigure}
    \caption{Experimental analysis of random sampling of clients and clusters.}
    \label{fig:cluster_plot2}
\end{figure*}

Considering $K=20$ and assuming that we have a total number of clients $n=6000$, we randomly assign each client to one of $20$ clusters. Then, we randomly sample some clients and make the observations depicted in Figure ~\ref{fig:cluster_plot2}.
Figure \ref{sfig:a2} shows that after sampling some clients randomly at uniform, the average number of clients from each cluster is almost the same, i.e., the number of samples lies within $1054$ to $1044$. Further, Figure ~\ref{sfig:b2} shows that the probability of getting $> 15\%$ error in the preceding experiment is as small as $7.5\%$ when averaged over 500 runs. Finally, Figure ~\ref{sfig:c2} shows that the minimum number of samples in each cluster is also the same. Hence, these three experiments support our  Lemma~\ref{Lemma:1}.

%\sandy{Grammar in following lemma makes it hard to follow. Please verify my edits.}
%%%%%%%%%% LEMMA 2 %%%%%%%%%%%
\begin{lemma}
\label{Lemma:2}
If~$\widebar{V}_i^\tau = \frac{1}{n\tau} \sum_{i\in C^\tau} V_i$ denotes the average of item vectors over some $C^\tau$ clients and $\widebar{V}_i^n = \frac{1}{n} \sum_{i=1}^{n} V_i$ represents the average of item vectors over a total of $n$ clients, then  $\mathbb{E}[U_u^T \widebar{V}_i^\tau] = \mathbb{E}[U_u^T \widebar{V}_i^n]$.
\end{lemma}
Here, $U_{u}^{T}\widebar{V}_{i}^{\tau}$ and $U_u^T\widebar{V}_i^n$ denote the predicted rating of a user $u$ obtained by using item gradients from sampled clients and all clients,  respectively. %This lemma holds inherently true if the underlying clients are homogeneous, which is not true in FL. Thus 
\\
\textbf{Proof. }Let $S_j$ denote the set of clients in cluster $j$ and $C_j$ denote the set of clients sampled from cluster $j$. Since in each cluster, the item vectors come from an identical distribution, and we have $\mathbb{E}[U_u^T\bar{V}_i^{S_i}] = \mathbb{E}[U_u^T\bar{V}_i^{C_i}]$. Here, $\bar{V}_i^{S_i}, \bar{V}_i^{C_i}$ represent the average of item vectors $V_i$'s from sets $S_i$ and $C_i$,  respectively. From Lemma ~\ref{Lemma:1}, if $\tau$ fraction of clients are selected from $n$ clients uniformly at random without repetition, then we have $C_j \approx \tau S_j$ with high probability. Then,\\
$$\mathbb{E}[U_u^T\bar{V}_i^\tau] = \frac{1}{\sum_{j=1}^l|C_i|}\sum_{j=1}^l |C_i|\mathbb{E}[U_u^T\bar{V}_i^{C_i}]$$
    $$= \frac{1}{\sum_{j=1}^l\tau|S_i|}\sum_{j=1}^l \tau|S_i|\mathbb{E}[U_u^T\bar{V}_i^{S_i}]$$
$$    =\mathbb{E}[U_u^T\bar{V}_i^n].
$$
We use the latent clustering assumption and Lemma \ref{Lemma:1} to prove that the expected values of predicted ratings of sampled and the entire population are equal. We now state our main theorem.
%%%%%% THEOREM %%%%%%%%
\begin{theorem}[\textbf{Random Sampling of Clients}]
\label{thm:main}
Given a rating matrix $R$, let $<\{U_u\}_{u=1}^{n}, \{V_i\}_{i=1}^{m}>$ denote a federated recommendation model with predicted ratings lying within a range of $[a, b]$. If $\overline{V}_i^\tau = \frac{1}{n\tau} \sum_{i \in C^{\tau}} V_i$ and $\widebar{V}_i^n = \frac{1}{n} \sum_{i=1}^{n} V_i$ represent the average of item vectors over some $\tau$ fraction of clients and total $n$ clients, respectively, then\\ 
$$\mathbb{P}(|U_u^T\widebar{V}_i^\tau - U_u^T\widebar{V}_i^n| \ge \epsilon) \le 2\exp\bigg\{\frac{-n\tau\epsilon^2}{2(b - a)^2}\bigg\}.$$
\end{theorem}
%%%%%% THEOREM %%%%%%%%
The preceding theorem can be proved using Hoeffding's bound and Lemma~\ref{Lemma:2}.\\
\noindent\textbf{Proof.}
From Hoeffding's inequality, we have \\
$\mathbb{P}(|U_u^T\bar{V}_i^\tau - \mathbb{E}[U_u^T\bar{V}_i^\tau]| \ge \epsilon) \le 2\exp\{\frac{-2n\tau\epsilon^2}{(b - a)^2}\}$, and\\
$\mathbb{P}(|U_u^T\bar{V}_i^n - \mathbb{E}[U_u^T\bar{V}_i^n]| \ge \epsilon) \le 2\exp\{\frac{-2n\epsilon^2}{(b - a)^2}\}.$

Thus, with probability at least $1-2\exp\{\frac{-2n\tau\epsilon^2}{(b - a)^2}\}$, we have 
\begin{align*}
    &\mathbb{E}[U_u^T\bar{V}_i^\tau] - \epsilon \le U_u^T\bar{V}_i^\tau \le \mathbb{E}[U_u^T\bar{V}_i^\tau] + \epsilon
\end{align*}    
\begin{align*}
    \implies \mathbb{E}[U_u^T\bar{V}_i^n - \epsilon \le U_u^T\bar{V}_i^\tau \le \mathbb{E}[U_u^T\bar{V}_i^n] + \epsilon \\\tag*{(From Lemma \ref{Lemma:2})}
\end{align*} 
\begin{align*}
    &\implies U_u^T\bar{V}_i^n - 2\epsilon \le U_u^T\bar{V}_i^\tau \le U_u^T\bar{V}_i^n + 2\epsilon.
\end{align*}
Thus, we get $\mathbb{P}(|U_u^T\bar{V}_i^\tau - U_u^T\bar{V}_i^n| \ge \epsilon) \le 2\exp\{\frac{-n\tau\epsilon^2}{2(b - a)^2}\}$.

According to the preceding theorem, if the ratings lie between $[1, 5]$, then the probability is less than $1\%$ that the error in the predicted rating exceeds $10\%$, which can be obtained with just $35\%$ of the clients from the pool of $6000$ clients. Therefore, from our theorem, if a dataset has around $6000$ clients, we choose $\tau = 0.35$ in our experiments.

While bounding sample complexity is generally difficult, the latent clustering of users makes it possible to provide non-trivial bounds. Without this assumption, straightforward use of Hoeffding's inequality would give trivial bounds of $100\%$ clients on the sample complexity, whereas we need only $35\%$. Importantly, this assumption is used only for the theoretical analysis of random sampling and does not affect the fairness of \ouralgo.  Without a uniform distribution assumption, getting non-trivial bounds is not possible. To instantiate with two clusters, one having $1000$ users and the other having just $10$ users, it is difficult to acquire equal representation from each cluster, even with many samples, e.g., if we sample more than 20 users, cluster 1 would have more users. Even if we sample only $10$ users, the chance of both clusters having users within $10\%$ of each other is less than $54\%$. This worsens for additional clusters with non-uniform users. 

While we provided non-trivial bounds by making a  uniform distribution assumption, the class imbalance in such sampled clients can lead to biased outcomes. We thus introduce a fairness-aware strategy to mitigate the issue of demographic bias in FedRec based on random sampling, as discussed below.
%%%%%%%%%%%%%%%%%%%%%%%%%%%%%%%%%%%
\begin{figure*}[t!]
         \includegraphics[scale=0.50]{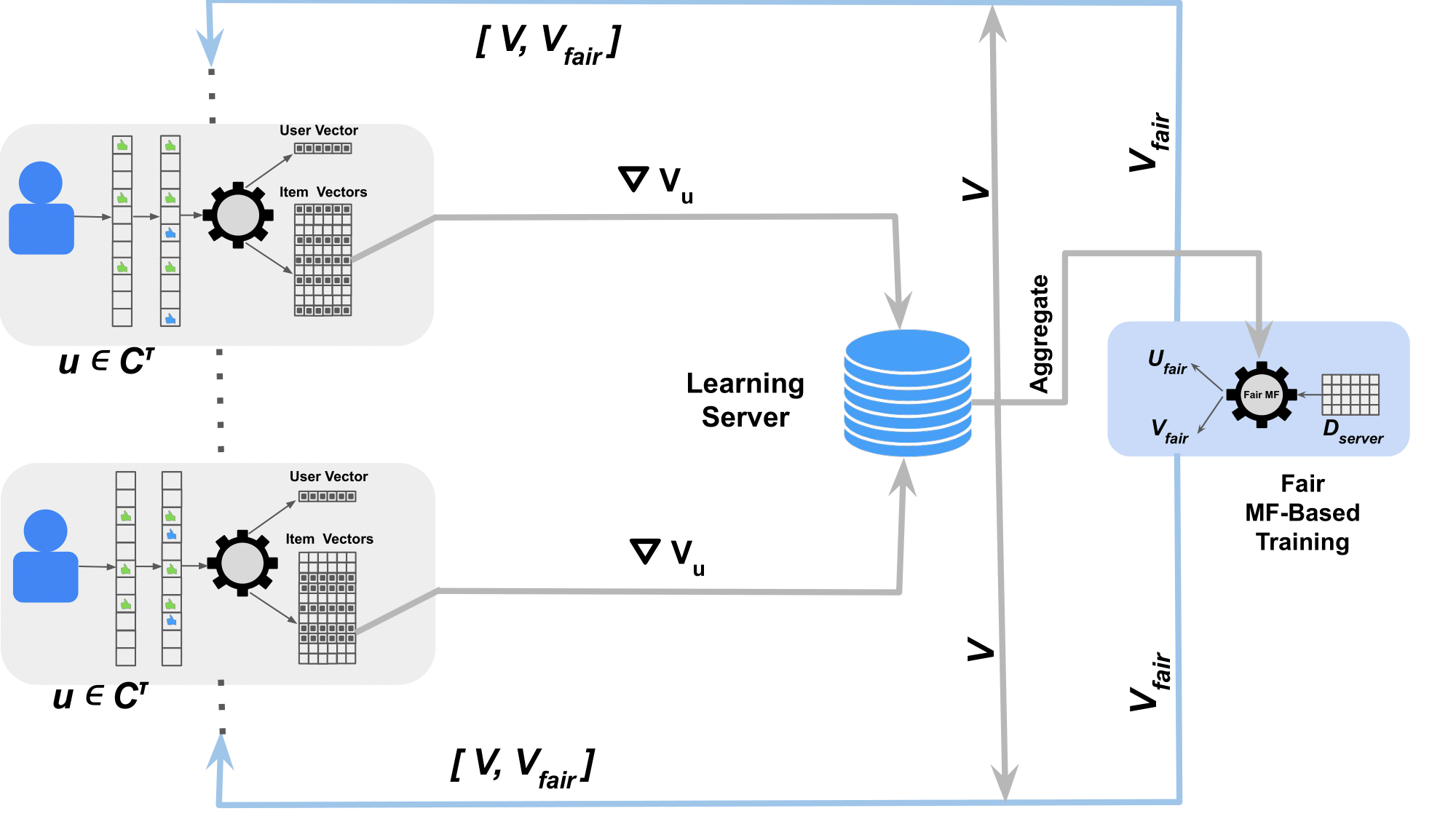}
         \caption{\textbf{The\ouralgo~Framework.}Every user trains a local recommendation model and generates user and item vectors. A fraction of clients $C^{\tau}$ are randomly sampled, and their item gradients are communicated to the server. The server aggregates these gradients for each item to generate global model $V$ and trains them towards fairness using FairMF to generate $U_{fair}$ and $V_{fair}$. Then, $V$ and $V_{fair}$ are communicated to every client. Each client trains $V$ to minimize the local loss and $V_{fair}$ to reduce the difference between the local and the fair global model. }
         \label{fig:mainfig}
\end{figure*}
%%%%%%%%%%%%%%%%%%%%%%%%%%%%%%%%%%%%
\begin{algorithm}[t!]
\caption{RS-FairFedRec Communication-Efficient and Fairness-Aware Federated Recommender System }\label{alg:ffedrec}
\hspace{-4mm}\textbf{Input:} $D_{server},D_{train},\tau,\gamma,\lambda^r,\lambda^f,\alpha,\rho$;
\textbf{Output.} $U_{u},V_{i}$ 
\begin{algorithmic}[1]
\State Initialize $V = [V_{i}]_{i=1}^m$ and $U_{fair} = [U_{u}]_{u=1}^{s}$
    \For{ $t= 1,2,......,T$}
        \State$C^{\tau} \leftarrow $ \{Randomly sub-sample $\tau$ fraction of clients from all the users in $D_{train}$\} 
        \State $ U_{fair} , V_{fair} \leftarrow$ FairMF($D_{server}$, $V$ , $U_{fair}$)
        %\State Communicate $V_{fair}$ to each client $u \in U$
        \For{each client $u \in U$ in  parallel}
            \State Sample $I^{'}_u$ from $I$ \textbackslash $ I_u$ with $| I^{'}_{u}| = \rho |I_u|$
            \State ClientFilling($V_{i}, i = 1,2,....,m; U_{u}; u ,t$)
            \State Compute $\nabla U_{u}$ $\&$ $\nabla V(u,i)$ by differentiating~\eqref{eq:local_FedRec}%via \eqref{eq:FedRec_usergradr}
            \State Update $U_{u}$ via $U_{u} \leftarrow U_{u} - \gamma \nabla U_{u} $
        \EndFor
            \For{$i = 1,.....,m$}
               \State Calculate $\nabla V_{i}$ for clients in $C^{\tau}$
                \State Update $V_i = V_i - \gamma \nabla V_{i}$
            \EndFor
        \State Decrease the learning rate $\gamma \leftarrow 0.9 \gamma$
    
    \EndFor
\end{algorithmic}
\end{algorithm}
\subsection{Dual-Fair Updating}
\label{sec:fair-mf}
%%%%%%%%%%%%%%%%%%%%%%%%%%%%%%%%%%%%
To mitigate the issue of demographic bias in sampling-based FRS, we propose a \textit{dual-fair update process, }which trains toward fairness in two separate phases: (1) FairMF, fairness-oriented global training, and (2) FO-Client Batch, local training to achieve fair item vectors. 
%%%%%%%%%%%%%%%%%%%%%%%%%%%%%%%%%%%%%
\subsubsection{\textbf{FairMF}}
We use fair matrix factorization (FairMF) to obtain a fair global model at the server. Let each client at the server be represented as $s \in \{1,....,S\}$ such that $S<<n$. FairMF (Algorithm \ref{alg:fairmf}) trains the data at server $D_{server}$ to obtain the global fairness objective by optimizing the loss function, which combines regularized MF in Equation \eqref{eq:MF_optimfactorization_func} and the fairness penalty in Equation \eqref{eq:d_bias}; this is given as
\begin{equation} 
\underset{U,V}{min} \; \mathcal{L}^{MF} + \lambda^{f}\mathcal{L}^{ap}
\label{eq:fairmf}.
\end{equation}
The hyperparameter $\lambda^{f}$ acts as a fairness penalizer. The update equations for $U_{u}$ and $V_{i}$ are obtained by taking derivatives of the fairness loss function with respect to client and item vectors, respectively. The server runs FairMF for some iterations $t_{s}$ and obtains $U_{fair}$ and $V_{fair}$. %Finally, $V_{fair}$ and $[V_{i}]_{i=1}^{m}$ (aggregated item vectors) are communicated to all the clients. 
%We provide the exact procedure for FairMF in .

\subsubsection{\textbf{FO-ClientBatch}} The fair global model is communicated to each local user. Clients in FedRec train locally using MF, as a result of which local item vectors can induce bias, rebiasing the overall global model. Thus, \emph{Fairness Oriented Client Batch} (FO-ClientBatch) ensures that clients update item vectors that are closest to the fair global model. Each client downloads both $V_{fair}$ and $V_i$ from the server. While $V_{fair}$ contributes towards fairness, communication of aggregated $V_i$ provides each user with the benefit of other users' participation. Since global fairness does not ensure local fairness in FL models, it is important that each local client also trains toward the fair model. This leads to the following local objective function:%Then, each user vector is updated locally, followed by an updation in the $V_{i}$ towards $V_{fair}$. Thus the local objective function changes to 
\begin{equation}
\underset{U,V}{min}\; \mathcal{L}^{MF} + \eta(||V_{fair}-V_i||^2)
\label{eq:local_FedRec}.   
\end{equation}
Clients keep $\nabla U_u$ and communicate only $\nabla V_i$. %The next section describes the detailed algorithm and communication details.
%-----------------------------%
%\begin{minipage}{0.47\textwidth}
\begin{algorithm}[t!]
    \centering
    \caption{FairMF ($D_{server}, U_{fair},V$)}\label{alg:fairmf}
    \begin{algorithmic}[1]
        \For{$T_{s} = 1,2,....,t_{s}$}
    \For{each  $(s,i)$ in $D_{server}$}
        \State if $r_{ui} \neq 0$, update $U_{s}$ and $V_{i}$ by calculating gradients by differentiating equation \eqref{eq:fairmf}
    \EndFor
\EndFor
\State$U_{fair} \leftarrow U_{s}$, $V_{fair} \leftarrow V_{i}$
\State \Return $U_{fair}, V_{fair}$
    \end{algorithmic}
\end{algorithm}
%\end{minipage}
%\hfill
%\begin{minipage}{0.5\textwidth}
\begin{algorithm}[t!]
    \centering
    \caption{ClientFilling( $V_i , i = 1,2,....,m ; U_{u}; u; t$)}\label{alg:clientfilling}
    \begin{algorithmic}[1]
    \If{ strategy == Hybrid Filling}
        \For{ $t_{local} = 1,2,.....,T_{local}$ }
            \State Calculate the gradient $\nabla U_{u}$
            \State  $U_{u} \leftarrow U_{u}- \gamma \nabla U_{u} $
        \EndFor
        \State Assign $r^{'}_{ui}$ to each item $i \in I^{'}_{u}$ via Hybrid Filling    
    \Else
        \State Assign $r^{'}_{ui}$ to each item $i \in I^{'}_{u}$ via User Averaging
        \EndIf
    \end{algorithmic}
\end{algorithm}
%\end{minipage}
%%%%%%%%%%%%%%%%%%%%%%%%

\subsection{\ouralgo}
\label{ssec:ouralgo}
%%%%%%%%%%%%%%%%%%%%%%%%
Algorithm \ref{alg:ffedrec} describes \ouralgo. We provide an overview of the proposed framework in Figure~\ref{fig:mainfig}. With the assumption of the availability of an extremely small dataset ($D_{server}$) corresponding to $S$ $(S << n)$ users at the server, 
%*********Initialization*****************
the procedure begins by randomly initializing $V = [V_{i}]_{i=1}^{m}$ and $U = [U_{fair}]_{u=1}^{S}$, which represent the item and user vectors, respectively. For the initial round, the server communicates only $V$ with every client for local training.
%***********Local DP and sampling *****************
 Locally, each client rates some items, and FedRec uses user averaging and hybrid filling to improve user privacy. In user averaging, $\rho$ unrated items are sampled, and then a virtual rating $r^{'}_{ui}$ is assigned to these items. We use the average rating of the observed items as the virtual rating. In hybrid filling, after a certain amount of time $T_{predict}$, the virtual rating is replaced by the predicted rating. To acquire a predicted rating, each client evaluates its user gradients and updates its user vector; this procedure is called \textit{client filling}~\cite{FedRec} (Algorithm \ref{alg:clientfilling}). The user and item gradients are evaluated simply by differentiating the \eqref{eq:local_FedRec}.  
 
Then, some ${\tau}$ fraction of users $C^{\tau}$ are randomly sampled, who communicate item updates to the server.
%************Server Aggregation*****************
 Notably, each user rates various items, and other items are unobserved. Since MF works by generating item vectors for each item, each user device generates a $(1 \times k)$ vector for one item, which is stacked with other item vectors to generate an item gradient matrix. At the server, the gradient corresponding to each item transmitted by $C^{\tau}$ users is aggregated using Equation~\ref{eq:FedRec_Itemvector}, and these gradients for each item are stacked to produce global $V$. 
%*************Server FairMF*********************
%\sandy{Following isn't clear. Do you mean V serves as initial item vectors? If so, can you rewrite to clarify? E.g., "V, initial item vectors, is further fed to FairMF"?} \kiran{done}
$V$, the aggregated global model is trained towards fairness by using Using $D_{server}$. The fair training at the server generates $V_{fair}$ and $U_{fair}$ using FairMF.
Along with $V_{fair}$, aggregated $V$ is also sent to each client to retain the federated properties of FedRec and allow benefits from the participation of other clients.
%*******Local trianing**************

The fair item vectors $V_{fair}$ and aggregated item vectors $V$ communicated by the server are received by each client $u$, and local training occurs at all clients. $V$ is used as an item vector by each local client to minimize the loss function in Equation~\ref{eq:MF_optimfactorization_func}, and $V_{fair}$ is used to minimize the difference between locally learnt item vectors and globally fair item vectors, as depicted by Equation~\ref{eq:local_FedRec}.
%*******Client Filling**************
 The updated item gradients are uploaded to the server to obtain aggregated vectors, train FairMF on the $D_{server}$, and obtain $U_{fair}$ and $V_{fair}$. The process repeats until convergence.
 %%%%%%%%%%%%%%%%%%%%%%%%%
\begin{figure}[t!]
\centering
\begin{subfigure}[t]{0.233\textwidth}
    \centering
    \includegraphics[scale=0.13]{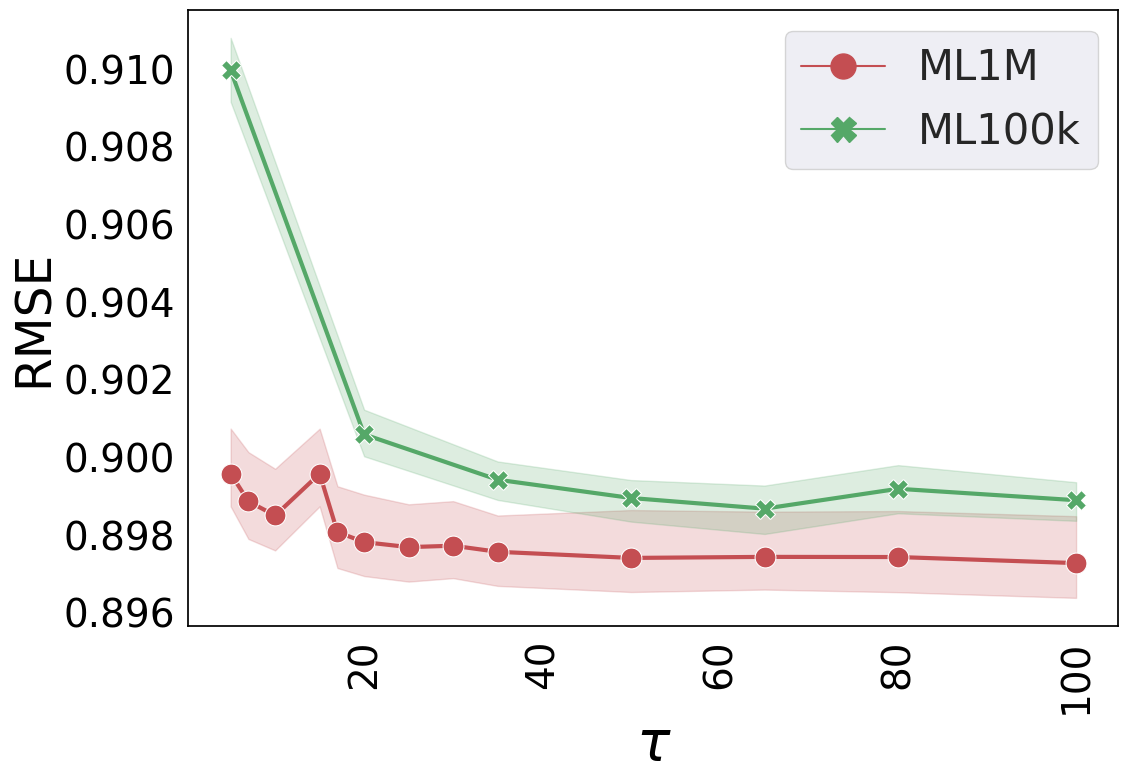}
    \caption{RMSE }
    \label{fig:Exp1_error}
\end{subfigure}
\hfill
\begin{subfigure}[t]{0.233\textwidth}
    \centering
    \includegraphics[scale=0.125]{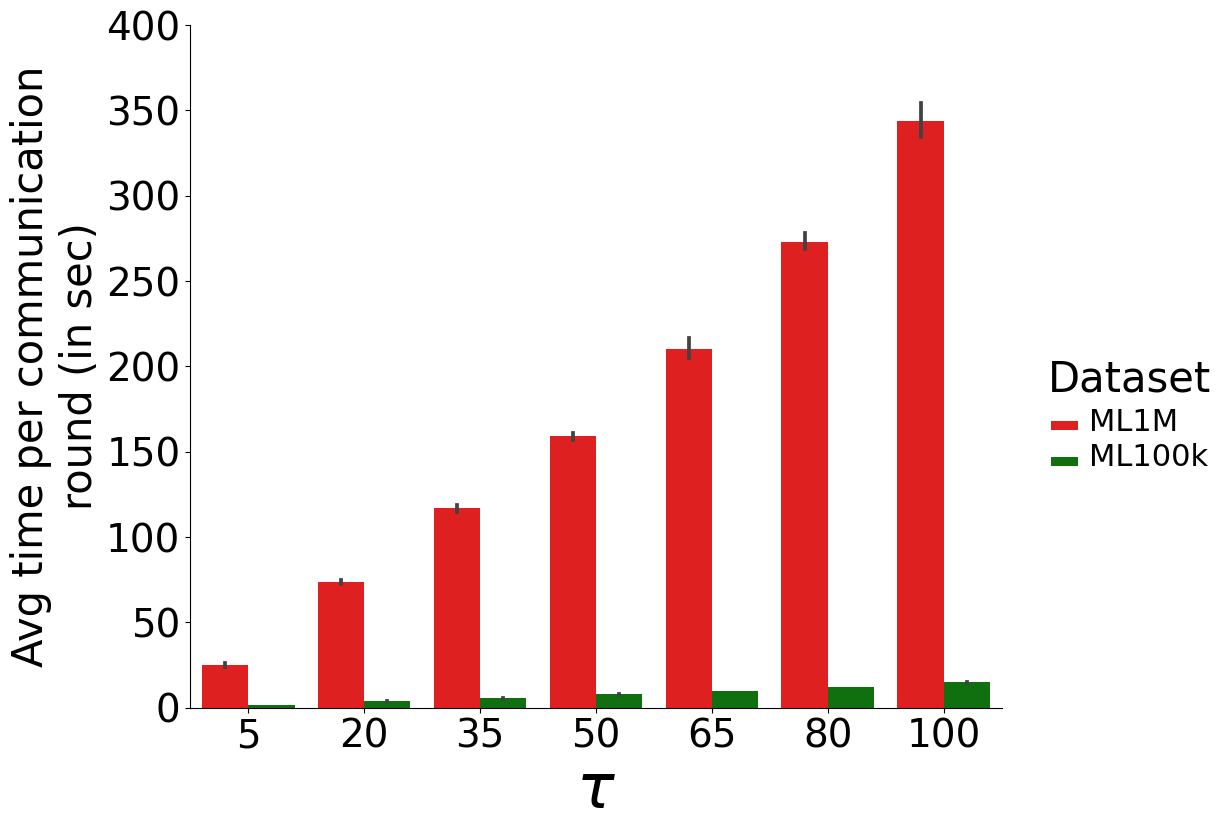}
    \caption{Time(in sec)}
    \label{fig:Exp1_Time}
\end{subfigure}
\caption{Comparison plots for accuracy and average time per communication round on different values of $\tau$ in RS-FedRec on two datasets, ML1M and ML100k.}
\end{figure}

\begin{figure}[t!]
     \centering
     \begin{subfigure}[t]{0.23\textwidth}
         \centering
         \includegraphics[width=\textwidth]{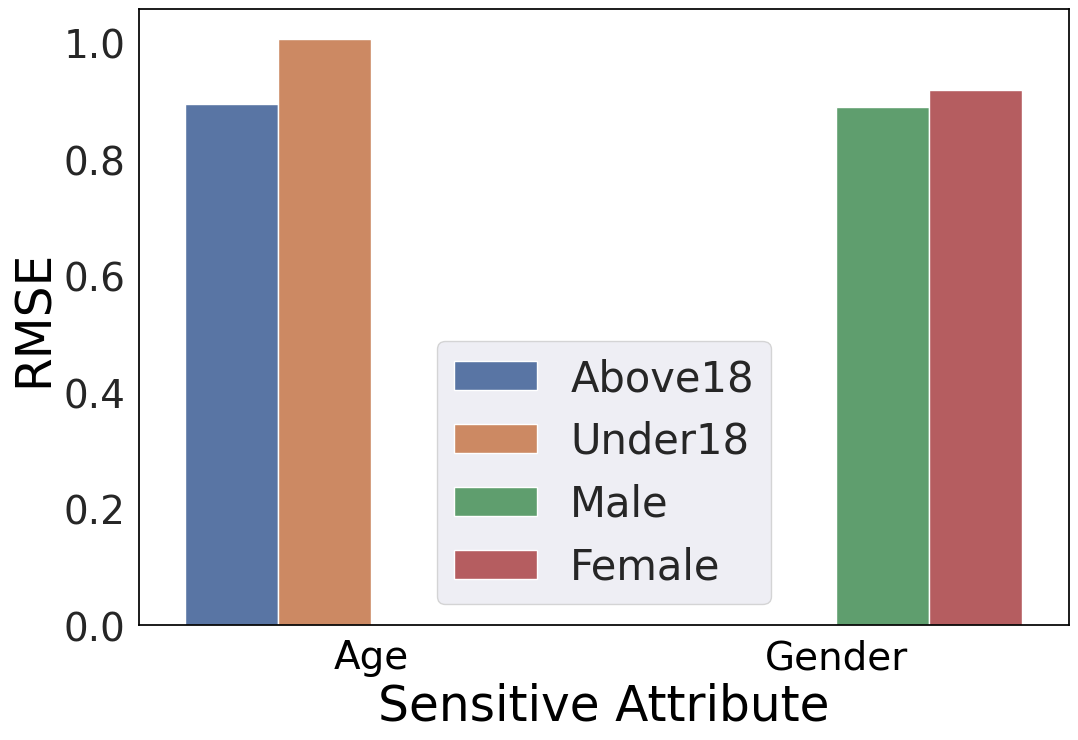}
         \caption{ML1M}
         \label{sfig:a_bias}
     \end{subfigure}
     \hfill
     \begin{subfigure}[t]{0.23\textwidth}
         \centering
         \includegraphics[width=\textwidth]{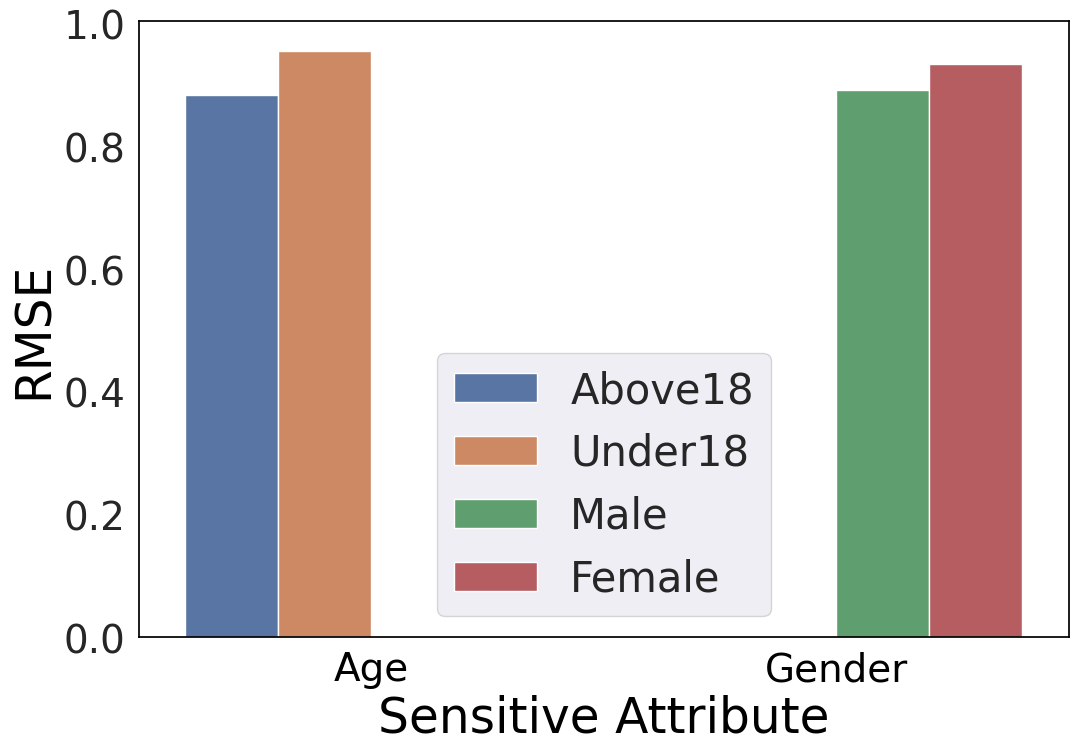}
         \caption{ML100k}
         \label{sfig:b_bias}
     \end{subfigure}
    
        \caption{Difference in FedRec RMSE scores for different sensitive attributes over two datasets.}
        \label{fig:Bias}
\end{figure}

%%%%%%%%%%%%%%%%%%%%%%%%%%%%%%%%%%%%%%%%%%%%%%%%%%%%%%%%%%%%

\begin{table*}[ht!]
\centering
\begin{tabular}{|c|c|c|c|c|c|c|c|}
\hline
\multirow{2}{*}{Dataset} & \multirow{2}{*}{$\#(Ratings)$} & \multirow{2}{*}{$\#(Movies)$} & \multirow{2}{*}{$\#(Users)$} & \multicolumn{2}{c|}{Gender} & \multicolumn{2}{c|}{Age Groups} \\ \cline{5-8}
                          &                           &                               &                         & $\#(Males)$        & $\#(Females)$       & $\#(\ge18)$        & $\#(<18)$        \\ \hline
ML1M                      & 1,000,209                 & 3,706                         & 6,040                   & 4,331        & 1,709          & 5,818        & 212          \\ \hline
ML100K                    & 100,000                   & 1,682                         & 943                     & 670          & 273            & 889          & 54           \\ \hline
\end{tabular}
\caption{Comparison of ML1M and ML100K datasets by gender and age.}
\label{tab:dataset_comparison}
\end{table*}

\begin{figure*}[t!]
     \centering
     \begin{subfigure}[b]{0.22\textwidth}
         \centering
         \includegraphics[width=\textwidth]{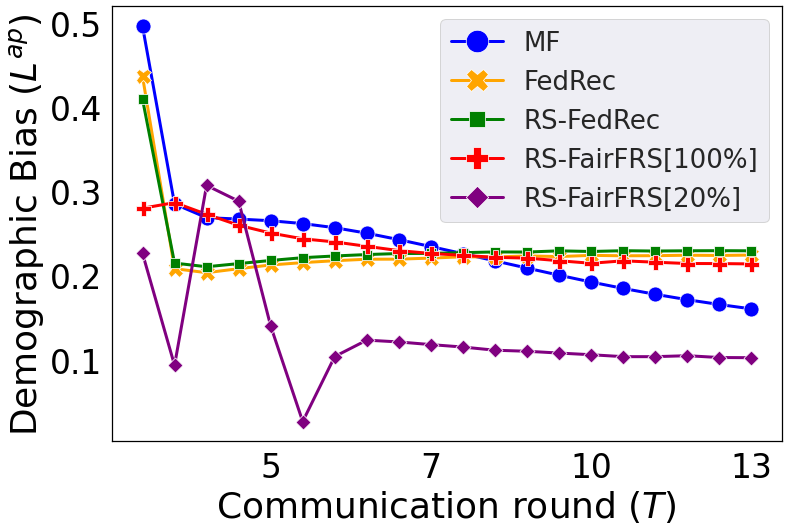}
         \caption{ML1M[Age.]}
         \label{sfig:a_}
     \end{subfigure}
     \hfill
     \begin{subfigure}[b]{0.22\textwidth}
         \centering
         \includegraphics[width=\textwidth]{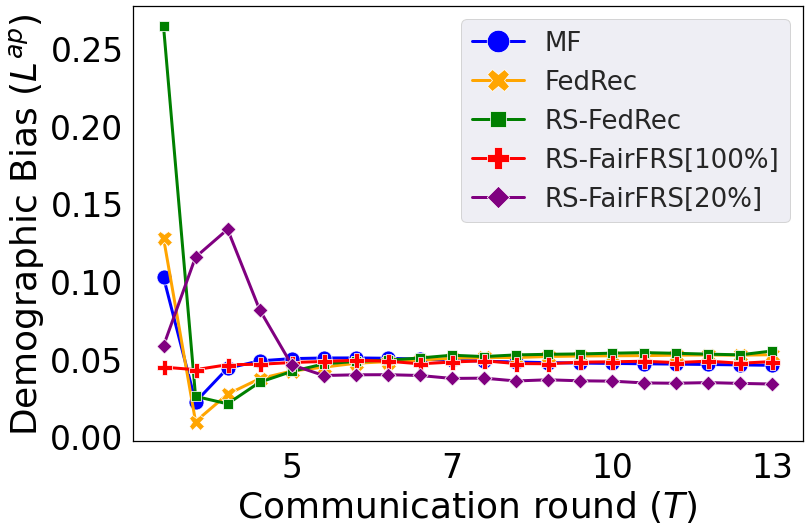}
         \caption{ML1M[Gender.]}
         \label{sfig:b_}
     \end{subfigure}
     \hfill
     \begin{subfigure}[b]{0.22\textwidth}
         \centering
         \includegraphics[width=\textwidth]{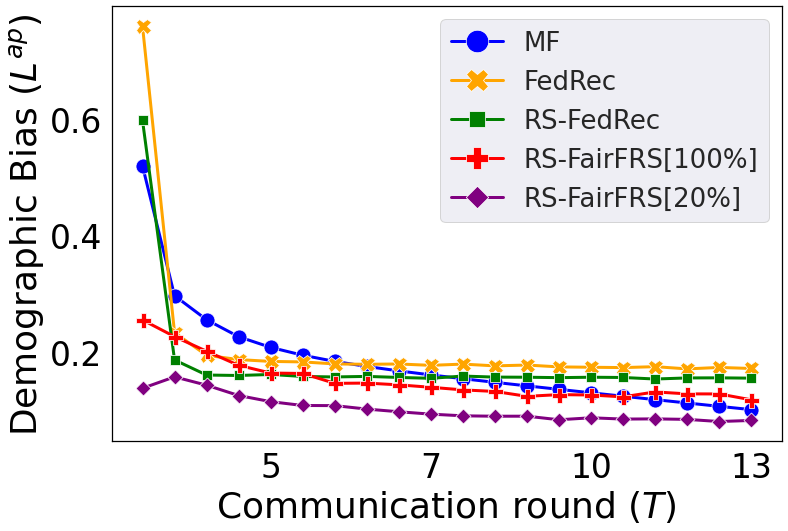}
         \caption{ML100k[Age.]}
         \label{sfig:c_}
     \end{subfigure}
          \hfill
     \begin{subfigure}[b]{0.22\textwidth}
         \centering
         \includegraphics[width=\textwidth]{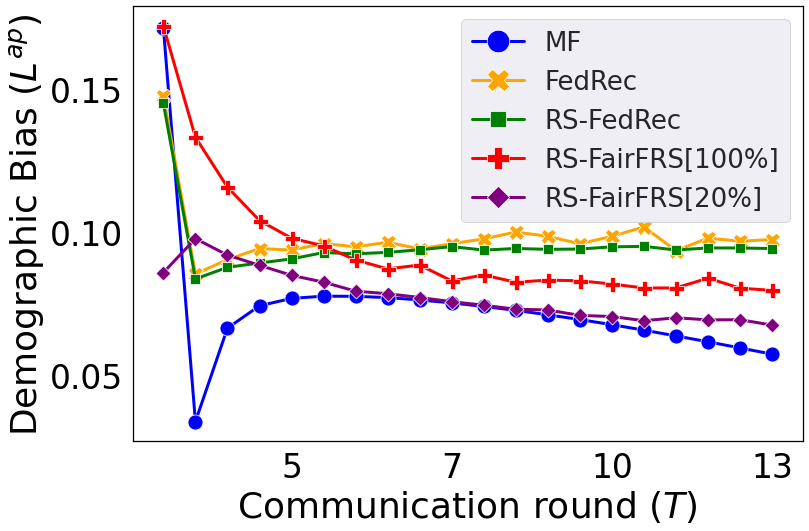}
         \caption{ML100k[Gender.]}
         \label{sfig:d_}
     \end{subfigure}

        \caption{ Comparison of $\mathcal{L}^{dap}$ on two datasets, ML1M and ML100k, for different algorithms. For each dataset, we show results for two different demographics, age and gender.}
        \label{fig:BiasTrends}
\end{figure*}

\begin{figure*}[t!]
     \centering
     \begin{subfigure}[t]{0.22\textwidth}
         \centering
         \includegraphics[width=\textwidth]{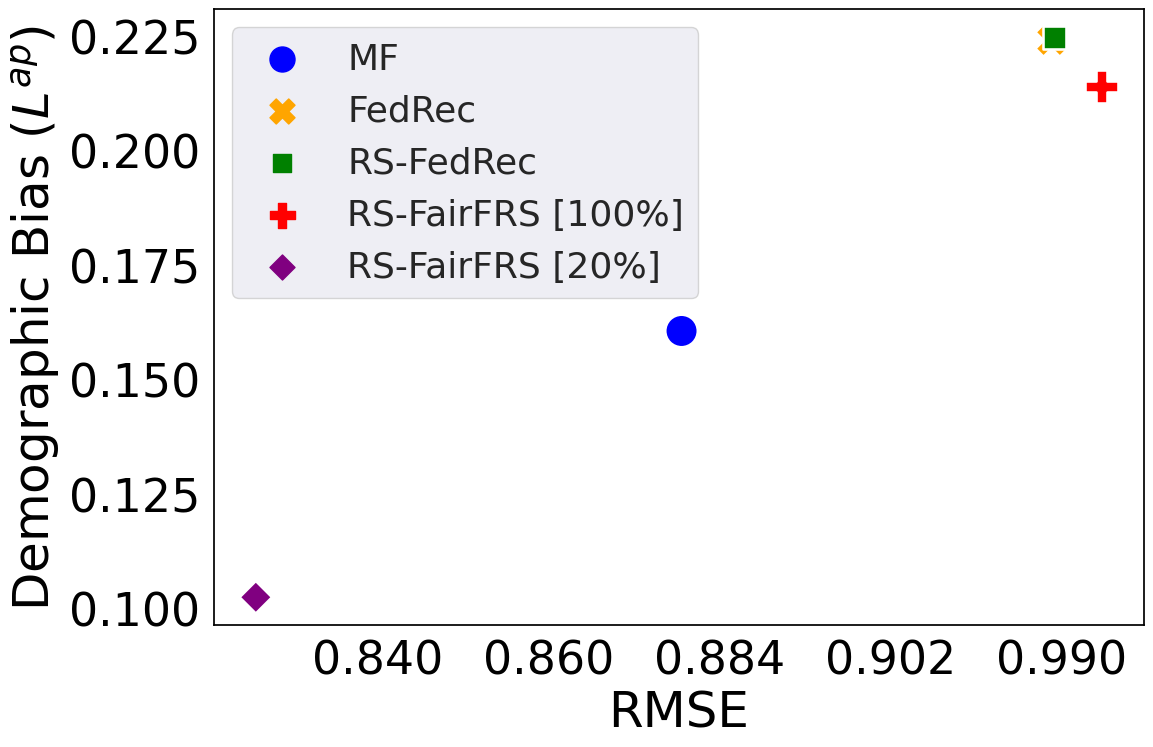}
         \caption{ML1M, Age.}
         \label{sfig:a}
     \end{subfigure}
     \hfill
     \begin{subfigure}[t]{0.22\textwidth}
         \centering
         \includegraphics[width=\textwidth]{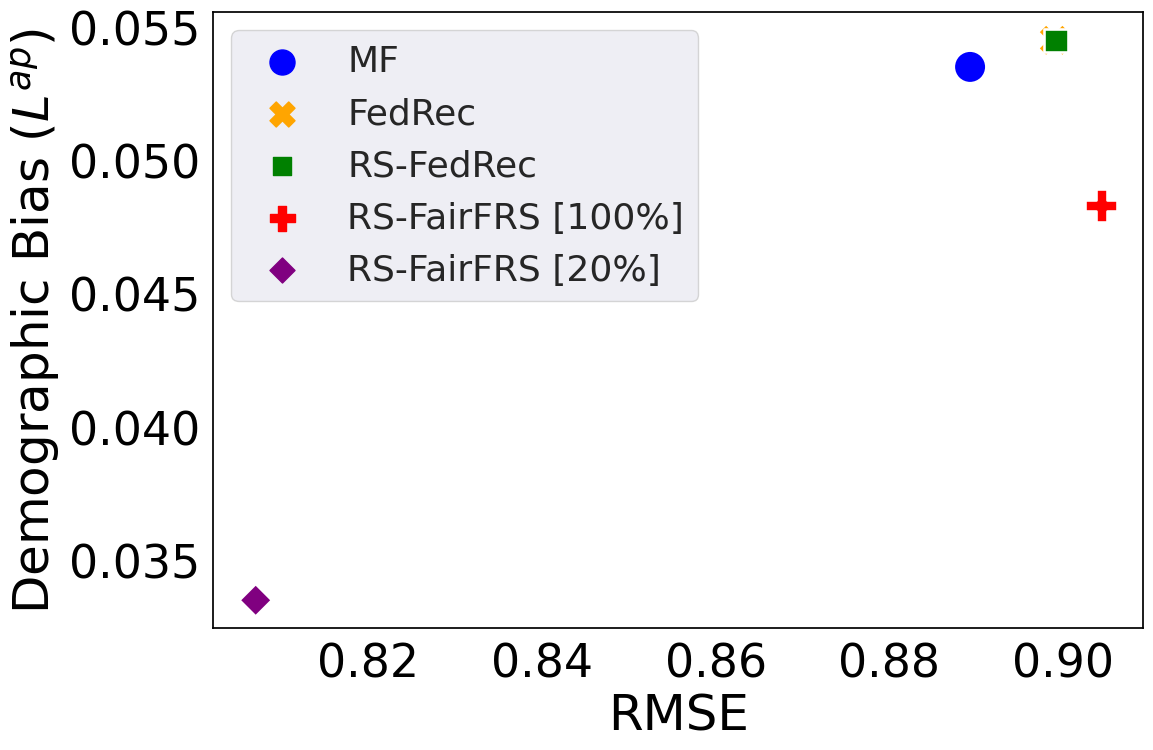}
         \caption{ML1M, Gender.}
         \label{sfig:b}
     \end{subfigure}
     \hfill
     \begin{subfigure}[t]{0.22\textwidth}
         \centering
         \includegraphics[width=\textwidth]{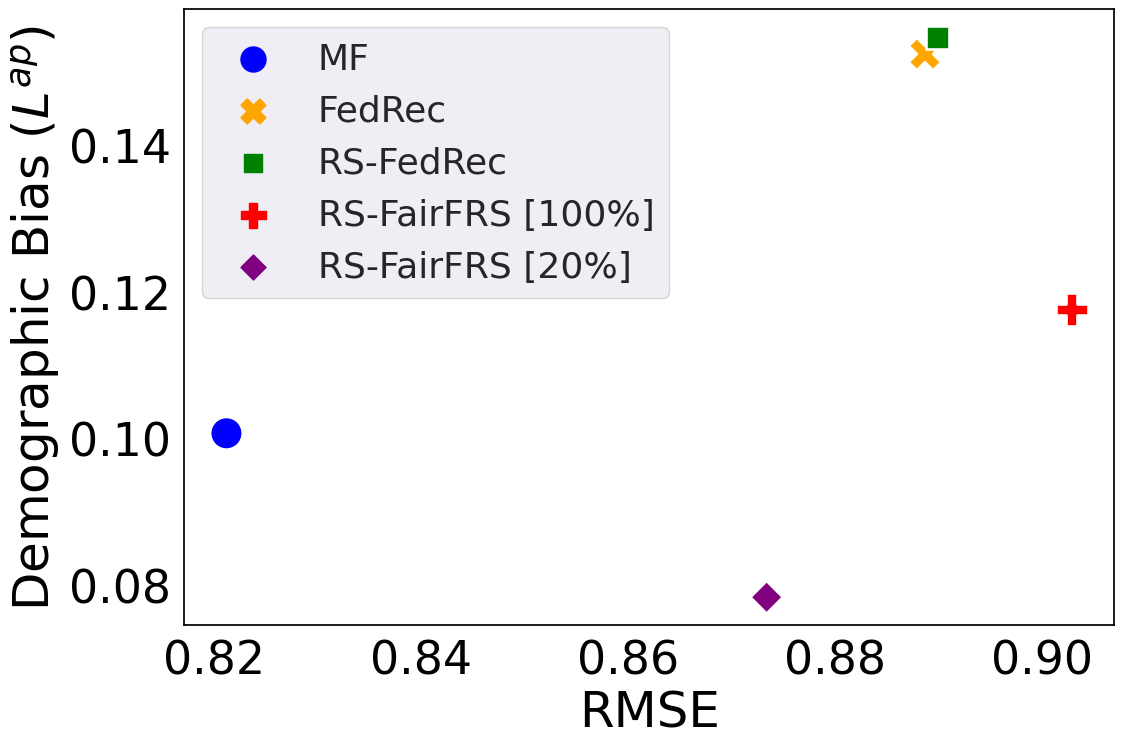}
         \caption{ML100k, Age.}
         \label{sfig:c}
     \end{subfigure}
    \hfill
     \begin{subfigure}[t]{0.22\textwidth}
         \centering
         \includegraphics[width=\textwidth]{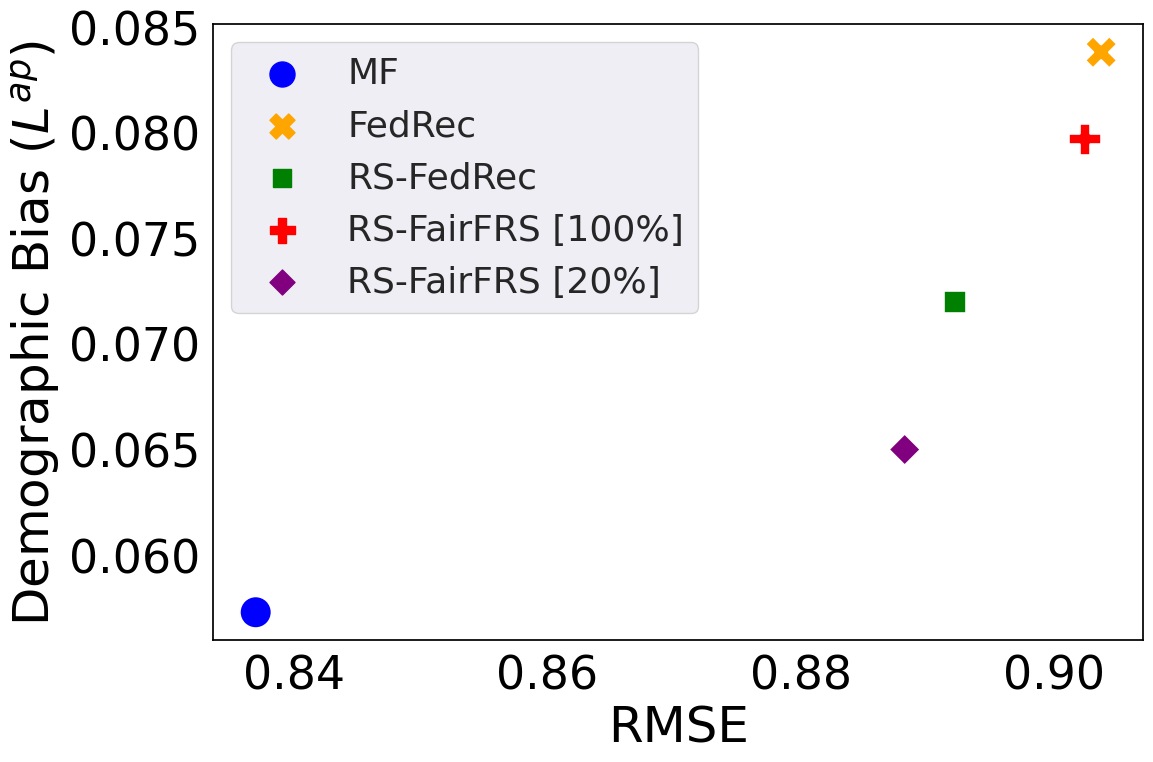}
         \caption{ML100k, Gender.}
         \label{sfig:d}
     \end{subfigure}
    \caption{Fairness vs accuracy plots for two datasets, ML1M and ML100k, for different algorithms. For each dataset, we show results for two different demographics, age and gender.}
    \label{fig:scatter_plot}
    \end{figure*}
\section{Experiments}
We designed experiments to validate and further elucidate the concepts and hypotheses we presented previously. Through empirical analysis, we rigorously examine and analyze these specific research questions: 
\begin{itemize}
    \item \textbf{RQ1}: What is the ideal fraction of clients that can be sampled in each communication round to reduce communication costs and retain model accuracy in FRS?

    \item \textbf{RQ2}: Can the final recommendations generated by FRS be biased against certain demographics? 

    \item \textbf{RQ3}: How effective is the proposed model \ouralgo~in reducing bias in federated scenarios? What is the least amount of data required at the server to generate a fair global model?

    \item \textbf{RQ4}: How does \ouralgo~affect model accuracy? Is there a trade-off between fairness and accuracy? 
\end{itemize}

\subsection{Experiment Setup}
\textbf{Datasets.  } We use two benchmark real-world datasets, \textit{ML1M} and \textit{ML100K}~\cite{Lens_2021}, since numerous past works on fair recommendations, including our baseline model FedRec, also use these datasets. Both datasets have explicit ratings ranging from 1 to 5. Table~\ref{tab:dataset_comparison} summarizes dataset statistics and illustrates the class imbalance. Since both datasets have users who have rated at least 5 movies, we consider entire datasets for our experiments. We split our datasets into $80:20$ ratio for training and testing, respectively. 

\textbf{Evaluation Protocol. } Similar to~\citet{FedRec}, we use $k=20$ latent features and set the value of the sampling parameter to $\rho = 2$. We run all the models to convergence $(T=20)$, and the FairMF procedure is executed to $t_s = 15$ runs. We report all results over an average of $10$ runs. We evaluate our model on the basis of (1) fairness and (2) accuracy. We use the metric $\mathcal{L}^{dap}$ to  measure fairness and RMSE (root mean square error) to measure accuracy. Importantly, our paper discusses the task of rating prediction and not the top-k ranking. Thus, we do not consider ranking metrics like NDCG, F1-score, etc. While it can be apparent that more data on the server will improve performance, we run all experiments on \ouralgo\ with $100\%$ and \ouralgo\ with $20\%$ data on the server to show that even with all data on the server, \ouralgo\ with $20\%$ data can improve accuracy and fairness. We also conduct additional experiments by restricting the number of items a user has rated on the server. 

%***************************************%
\subsection{Empirical Evaluation and Discussion}
\label{ssec:exp}
%***************************************%

%****************************************************
\subsubsection{\textbf{Addressing RQ1:} Random Sampling of Clients.} We randomly sample $\tau$ fraction of clients in each communication round, and only those clients communicate their item gradients with the server. We analyze the accuracy of FedRec for a wide range of $\tau$ values, as shown in Figure \ref{fig:Exp1_error}. Corresponding to each value of $\tau$ in FedRec (x-axis), we report the RMSE value on the y-axis. The figure shows that for both ML1M and ML100K, given a value of $\tau = 35$ ($35\%$ of the total clients), the model performs nearly the same as with $\tau = 100$ ($100\%$ of the clients). This answers our first research question since we find $\tau = 35$ to be the ideal number of clients that, when sampled, will not degrade the overall performance of FedRec. 

This finding also aligns with our theoretical analysis presented in Section~\ref{sec:random-sampling}. From Theorem~\ref{thm:main}, it can be shown that the predicted rating of each client with sampled item vectors will be within $5\%$ error with a probability of at least $0.96$. These findings align with~\citet{largecohort} since the training loss increases with an increasing sample size. 

We also present the cost analysis by plotting the average time consumed by each communication round against $\tau$ in Figure \ref{fig:Exp1_Time}. As we increase the fraction of clients sampled in each communication round, the time consumed increases, as well. From the two figures identified here, we conclude that the communication cost of FedRec can be reduced significantly (i.e., from 350 sec to 120 sec on average for each communication round) without affecting model accuracy by sampling $35\%$ clients in each round. Therefore, for all experiments in upcoming sections, we select $\tau = 0.35$ as the ideal value to provide reasonable accuracy in less time.

\subsubsection{\textbf{Addressing RQ2:} Disparate Treatment of FedRec}. We analyze the performance of FedRec on various sensitive attributes to demonstrate the biased treatment against certain groups. For this, we compute average losses on two binary sensitive attributes, age $( \ge 18 , <18)$ and gender (males and females) separately.  We observe that in ML1M, users less than 18 y.o. have an RMSE of $1.00784$, whereas those above $18$ y.o. have a lower RMSE of $0.8947$; female and male users experience an RMSE of $0.91902$ and $0.89112$, respectively. We observe a similar trend for the ML100K dataset, where RMSE for users above $18$ y.o. is $0.95684$ and below 18 y.o. is  0.88452. Females in the ML100K dataset acquire recommendations with $0.93465$, and males with $0.89249$ error. These results show disparate recommendations received by marginalized users.

Figure \ref{fig:Bias} shows that the groups with more clients \textit{(advantaged groups)} enjoy more accurate recommendations, whereas those with fewer clients \textit{(disadvantaged groups)} receive more erroneous recommendations. These experiments answer our second research question by showing that FedRec shows disparate treatment against underrepresented groups (females and users under 18) in both datasets. This shows that the existing class imbalance affects FedRec recommendations, which motivates our development of a fairer counterpart.

\begin{figure*}[th!]
     \centering
     \begin{subfigure}[t]{0.22\textwidth}
    \includegraphics[width=\textwidth]{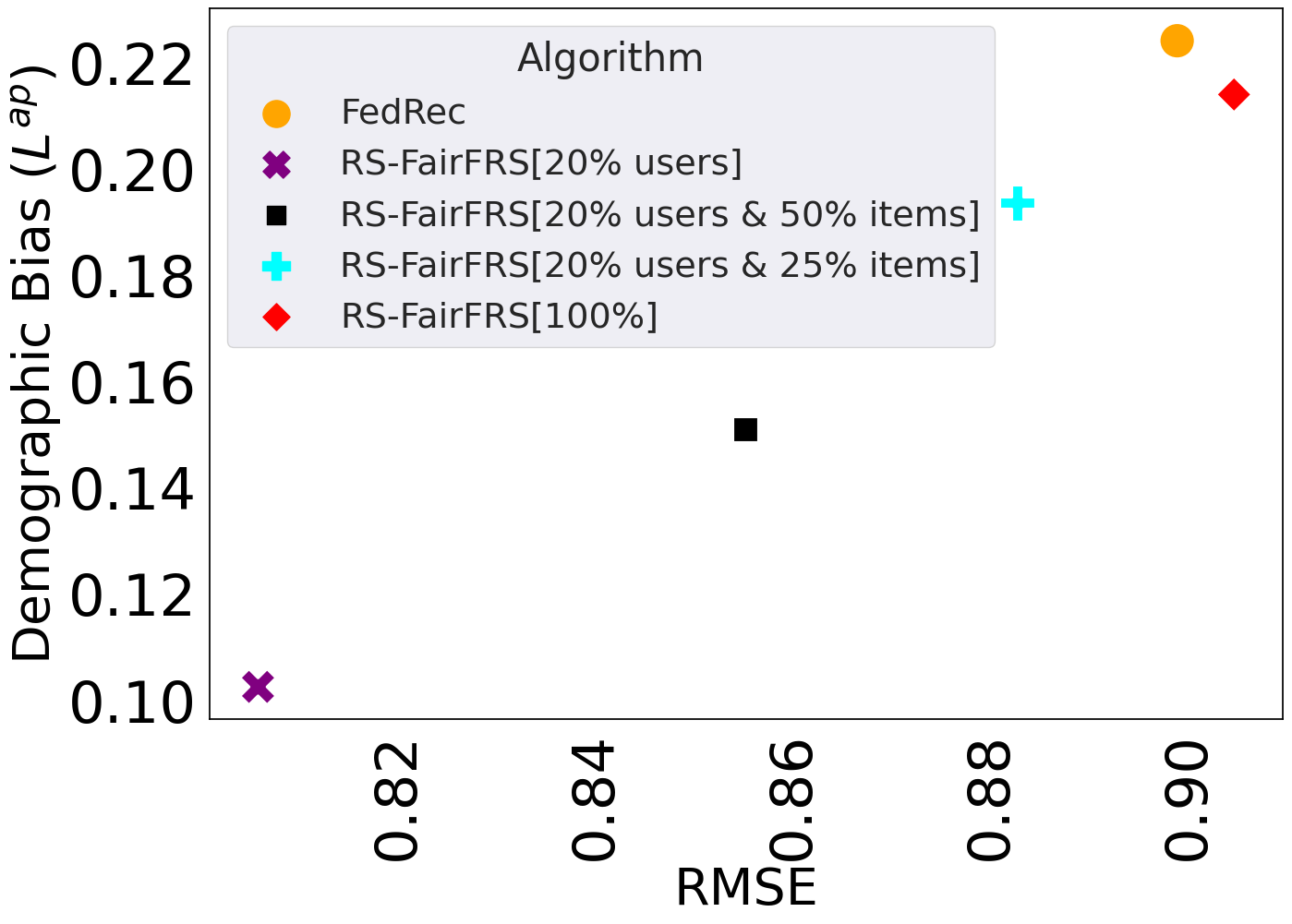}
         \caption{ML1M[Age.]}
         \label{sfig:ab_a}
     \end{subfigure}
     \hfill
     \begin{subfigure}[t]{0.22\textwidth}
     \includegraphics[width=\textwidth]{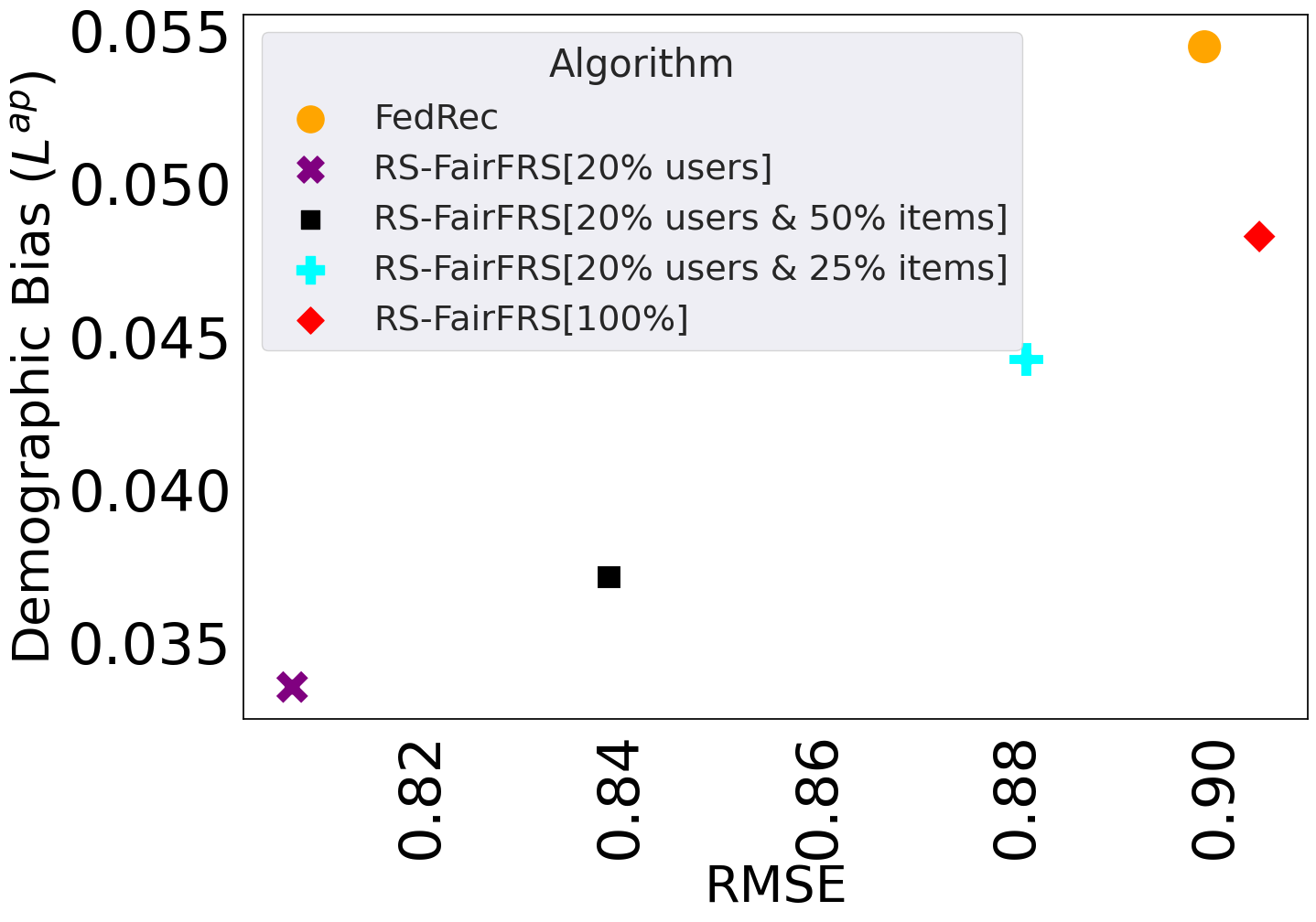}
         \caption{ML1M[Gender.]}
         \label{sfig:ab_b}
     \end{subfigure}
     \hfill
     \begin{subfigure}[t]{0.22\textwidth}
    \includegraphics[width=\textwidth]{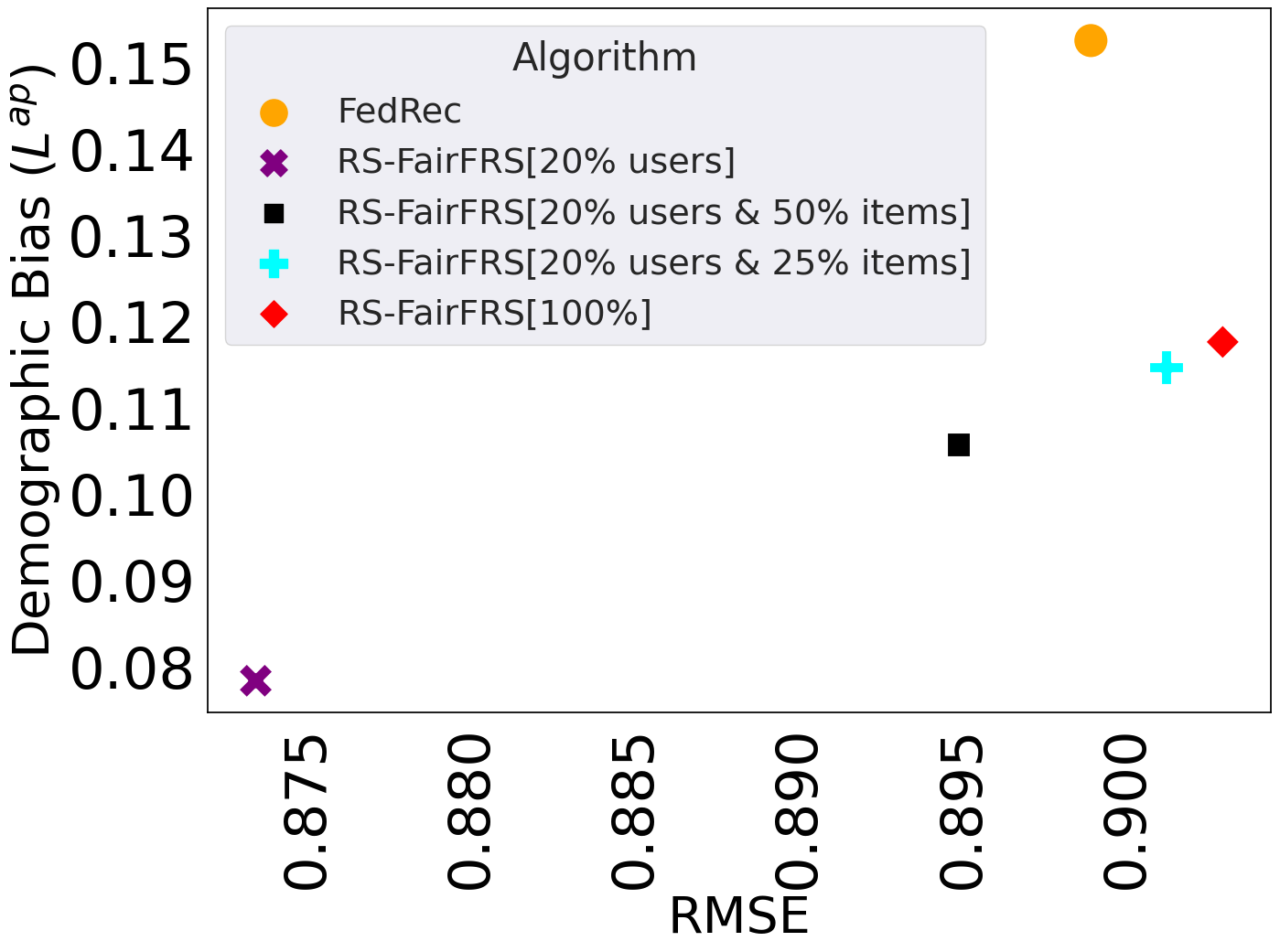}
         \caption{ML100k[Age.]}
         \label{sfig:ab_c}
     \end{subfigure}
    \hfill
     \begin{subfigure}[t]{0.22\textwidth}
    \includegraphics[width=\textwidth]{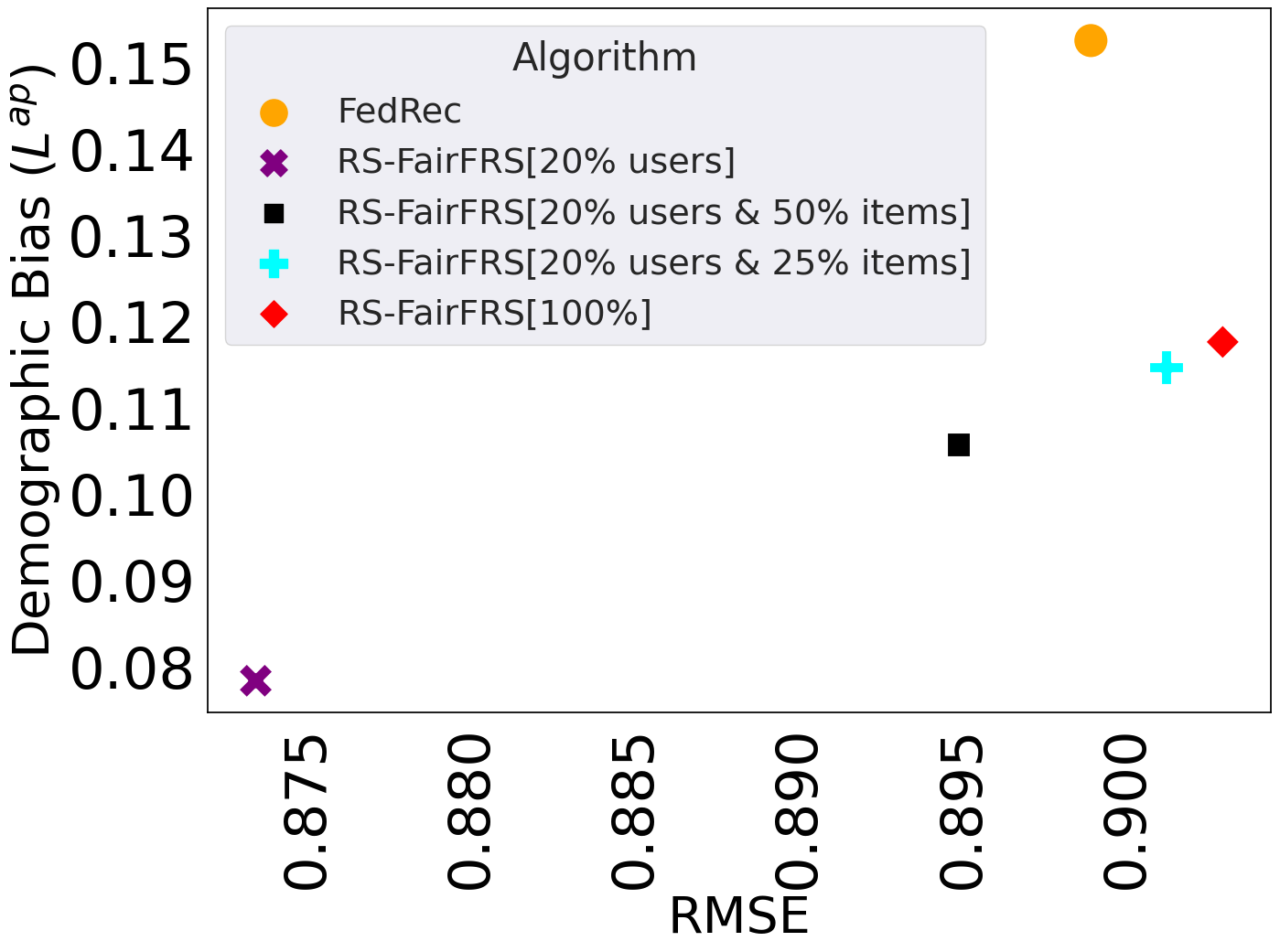}
         \caption{ML100k[Gender.]}
         \label{sfig:ab_d}
     \end{subfigure}
    \caption{Comparison of $\mathcal{L}^{dap}$ on two datasets, ML1M and ML100k, for \ouralgo with reduced items at the server. For each dataset, we show the results for two different demographics, age and gender.}
    \label{fig:ab_scatter_plot}
\end{figure*}
\subsubsection{\textbf{Addressing RQ3:} Fairness of \ouralgo.} To analyze the fairness of \ouralgo, we compare our results to Matrix Factorization, FedRec, RS-FedRec (FedRec with sampling), and \ouralgo\ with ($100\%$) and ($20\%$) data at the server. Figure~\ref{fig:BiasTrends} plots the demographic bias obtained using $\mathcal{L}^{dap}$ against communication rounds for all federated models and treats each epoch as a corresponding communication round for centralized MF. These results provide the following insights: (1) It is evident that the bias in FedRec significantly exceeds that in MF, which shows that FedRec amplifies bias due to averaging at the server. (2) Random sampling of clients does not noticeably affect the biased behavior of FedRec, which shows that sampling retains the overall performance of the model. (3) Seed data of size as small as $20\%$ of the entire amount of data among all participating clients can significantly reduce the demographic bias in FedRec; this cannot be achieved even by assuming $100\%$ of the data at the server. 

We note that having $100\%$ of the data at the server can be impractical. $20\%$ data on the server helps achieve greater fairness than $100\%$ because our model uses random sampling of clients in each round. $100\%$ data on the server includes many outliers, which worsens the fairness of sampled clients, thus degrading fairness.  Further, we observe that our model uses randomly initialized user and item vectors. The random vectors are trained and updated to obtain the closest rating predictions as the training proceeds toward minima. Thus, over some time, the curve smoothes out, but initial readings can fluctuate greatly. It is evident from all four graphs in Figure~\ref{fig:BiasTrends} that \ouralgo\ with $20\%$ data on the server acquires the least bias on both datasets for all sensitive attributes. Thus, \ouralgo $(20\%)$ significantly reduces the bias in FedRec without leaking any sensitive information of most users during training.

\subsubsection{\textbf{Addressing RQ4:} Fairness vs Accuracy in \ouralgo. } Many past works (~\cite{cooper2021emergent,lesota2022exploring,sonboli2022controlling,ge2022toward}) have researched the challenging issue of trade-offs between achieving fair vs accuracy in recommendations. In Figure \ref{fig:scatter_plot}, we plot $\mathcal{L}^{dap}$ against the RMSE value to depict the strong performance of our model in both aspects. We remark that a lower value in both demographic bias and  RMSE accounts for a 
%\sandy{Explain "better" below. Better in terms of what?}
better model. Users of both age groups in the ML100k dataset enjoy fair recommendations with \ouralgo$(20\%)$. In this dataset, MF performs slightly better than other algorithms for the gender attribute in terms of accuracy and fairness due to the efficiency of centralized settings for smaller datasets. Overall, however, \ouralgo$(20\%)$ outperforms all other federated algorithms. 

In addition to generating fairer recommendations, our model also yields the highest accuracy for two important reasons. First, the fairness-oriented training at the server uses Eq~\ref{eq:fairmf}, which incorporates loss $\mathcal{L}^{dap}$. This procedure trains to reduce the MSE loss along with reducing loss on both sensitive attributes. Second, the dual nature of local training of users incorporates Eq~\ref{eq:local_FedRec}. The clients initially reduce the loss between learnt and actual ratings and then reduce the difference between local and fair item vectors produced by FairMF. Thus, a careful selection of the values of $\eta$ and $\lambda^{f}$ affects the fairness vs accuracy trade-off.

\subsubsection{Additional Experiments. }Past experiments assume the availability of all rated items on the server. We toughen this consideration by using only $25\%$ and $50\%$ of previously rated items. Figure ~\ref{fig:ab_scatter_plot} presents the final results obtained on these additional experiments. We plot $\mathcal{L}^{dap}$ vs RMSE to represent our results using a scatter plot. This plot shows the promising results of our algorithm \textit{even when the dataset is further reduced on the server}. The consistency of results on accuracy and fairness are seen across all datasets and sensitive attributes. Though intuitive, our results suggest that having additional ratings can help to obtain fairer and more accurate results. However, with only $25\%$ of the item ratings, we can still significantly reduce bias and improve accuracy for all four datasets without sharing any sensitive attribute of active clients during federated training. 
%\sandy{Can you omit the following 2 sentences, which deflate your point?}
%Moving further, if we increase data on the server by considering $50\%$ of the item ratings, then our results are slightly better than the previous consideration of $25\%$. And the best outcomes are encountered with all ratings.

%\sandy{I have never seen these two sections combined. Let's just call it a Conclusion?}
%%%%%%%%%%%%%%%%%%%%%%%
\section{Conclusion and Future Work}
\label{sec:con}
%%%%%%%%%%%%%%%%%%%%%%%
We propose \ouralgo\, which incorporates the key ideas of a random sampling of $35\%$ of total clients to reduce the communication cost of FL systems and dual-fair updating to mitigate group bias. We are the first to theoretically bound the sample complexity of the fraction of clients needing to be sampled without affecting model accuracy. Further, we empirically demonstrate our theoretical results and provide experiments to show that the dual-fairness updation technique in \ouralgo\ can significantly reduce bias and improve model accuracy. 

Due to space constraints and distributed training expenses, we present results on FedRec and movie recommendation datasets. Furthermore, to the best of our knowledge, we are first to demonstrate the complexity of obtaining an ideal fraction of users that should participate in each round of communication in FRS. Though many communication-efficient techniques have been adopted in the past, client sampling remains indispensable to most federated models, as simultaneous communication by millions of users in practice can raise costs substantially.  Furthermore, many recommendation models can be adapted for federated training, we particularly tackle the issue of fairness in factorization-based models due to the inherent challenge that arises when only public features can be communicated with the server. We discussed how training local clients towards global vectors remains inefficient and thus develop dual-fairness updation strategy to mitigate the issue of fairness under unawareness of sensitive attributes in FRS in such complex models and empirically demonstrated the efficacy of our algorithm. However, any FR system using a factorization-based approach can adopt our framework, and we expect consistency in fairness and accuracy on any explicit recommendation dataset. We further aim to theoretically analyze the tradeoff between fairness and accuracy by analyzing both global and local regularization hyperparameters. We leave the analysis of \ouralgo on other algorithms and datasets to future work. Though this paper addresses demographic bias, FedRec faces various other biases caused by the heterogeneity of data, users and tasks (e.g., cross-domain recommender systems). We intend to explore other types of FRS biases as we continue to address this critical issue in the future.

%%
%% The next two lines define the bibliography style to be used, and
%% the bibliography file.
%\bibliographystyle{ACM-Reference-Format}
%\bibliography{acmart}

\bibliographystyle{ACM-Reference-Format}
\bibliography{sample-base}

\end{document}